\documentclass[10pt,journal,compsoc]{IEEEtran}

\ifCLASSOPTIONcompsoc
  \usepackage[nocompress]{cite}
\else
  \usepackage{cite}
\fi

\usepackage[english]{babel}
\usepackage{bm,graphicx,amssymb,url}

\begin{document}

\title{Lightweight Fault Tolerance in Large-Scale Distributed Graph Processing}

\author{Da~Yan$^1$,
        James~Cheng$^2$,
        Fan~Yang$^3$
\IEEEcompsocitemizethanks{\IEEEcompsocthanksitem The authors are with the Department of Computer Science and Engineering, The Chinese University of Hong Kong, Shatin N.T., Hong Kong.\protect\\
E-mails: \{$^1$yanda, $^2$jcheng, $^3$fyang\}@cse.cuhk.edu.hk}
}

\markboth{IEEE TRANSACTIONS ON KNOWLEDGE AND DATA ENGINEERING, Vol.~6, No.~1, January~2007}%
{Yan et al.: Lightweight Fault Tolerance in Pregel-Like Systems}

\IEEEtitleabstractindextext{%
\begin{abstract}
The success of Google's Pregel framework in distributed graph processing has inspired a surging interest in developing Pregel-like platforms featuring a user-friendly ``think like a vertex'' programming model. Existing Pregel-like systems support a fault tolerance mechanism called checkpointing, which periodically saves computation states as checkpoints to HDFS, so that when a failure happens, computation rolls back to the latest checkpoint. However, a checkpoint in existing systems stores a huge amount of data, including vertex states, edges, and messages sent by vertices, which significantly degrades the failure-free performance. Moreover, the high checkpointing cost prevents frequent checkpointing, and thus recovery has to replay all the computations from a state checkpointed some time ago.

\ \ \ \ \ \ In this paper, we propose a novel checkpointing approach which only stores vertex states and incremental edge updates to HDFS as a lightweight checkpoint (LWCP), so that writing an LWCP is typically tens of times faster than writing a conventional checkpoint. To recover from the latest LWCP, messages are generated from the vertex states, and graph topology is recovered by replaying incremental edge updates. We show how to realize lightweight checkpointing with minor modifications of the vertex-centric programming interface. We also apply the same idea to a recently-proposed log-based approach for fast recovery, to make it work efficiently in practice by significantly reducing the cost of garbage collection of logs. Extensive experiments on large real graphs verified the effectiveness of LWCP in improving both failure-free performance and the performance of recovery.
\end{abstract}

}

\maketitle

\IEEEdisplaynontitleabstractindextext
\IEEEpeerreviewmaketitle

\IEEEraisesectionheading{\section{Introduction}\label{sec:intro}}
\IEEEPARstart{S}{everal} Pregel-like systems have been developed recently for big graph analytics, such as Giraph~\cite{giraph}, GraphLab~\cite{graphlab,powergraph}, GPS~\cite{SalihogluW13ssdbm} and Pregel+~\cite{pregelplus}. These systems adopt a user-friendly vertex-centric programming model first proposed in Google's Pregel~\cite{pregel}, where a programmer only needs to specify the behavior of one generic vertex. Moreover, to be resilient to machine failures, these distributed systems support fault tolerance by checkpointing, which periodically saves the current state of computation to a failure-resilient storage such as HDFS\footnote{HDFS replicates each data block to three different machines so that data loss does not happen unless all the three machines crash at the same time.}. However, checkpointing incurs additional overhead during the failure-free execution, and one main goal of this paper is to significantly reduce this overhead by investigating the properties specific to the framework of Pregel.

\vspace{1.5mm}

\noindent{\bf Lightweight Checkpointing.} To explain why the conventional checkpointing method of Pregel is inefficient, we consider the problem of computing the PageRank of every vertex $v$ in a graph, denoted by $a(v)$. Each vertex $v$ also maintains the set of neighbors that it links to, denoted by $\Gamma(v)$. The Pregel job for PageRank computation proceeds in iterations called supersteps, and in each superstep, every vertex $v$ updates $a(v)$ by summing up the values sent from its in-neighbors in the previous superstep (and adjusting the sum by a damping factor); $v$ then distributes $a(v)$ evenly to its out-neighbors by sending each out-neighbor a message with value equal to $a(v)/|\Gamma(v)|$.

The above job for PageRank computation may be specified to save a checkpoint for every 10 supersteps. If a machine crashes, say, at superstep~17, then the latest checkpoint saved at superstep~10 will be loaded to roll the state of every vertex back to the end of superstep~10, and the computation then reruns from superstep~11.

A checkpoint written by an existing Pregel-like system is {\em heavyweight}, which saves the following data for every vertex $v$: (1)~value $a(v)$, (2)~adjacency list $\Gamma(v)$, and (3)~incoming messages to $v$ (for the next superstep), denoted by $M_{in}(v)$. We need to store $M_{in}(v)$ since it may be used to update $a(v)$ and to compute the messages that $v$ will send; we also need to store $\Gamma(v)$ since Pregel allows topology mutation, and $\Gamma(v)$ may change in different supersteps.

However, this solution is an overkill for PageRank computation, since it suffices to save a {\em lightweight} checkpoint, i.e., we save only the PageRank value $a(v)$ of every vertex $v$: (1)~outgoing messages of $v$ can be computed without examining $M_{in}(v)$ (recall that the message is $a(v)/|\Gamma(v)|$); (2)~$\Gamma(v)$ is static and can be directly loaded from the input graph.

We remark that while the lightweight checkpoint discussed above can be straightforwardly applied for PageRank computation in the framework of Pregel, this paper aims to apply lightweight checkpointing to a general Pregel algorithm. For this purpose, we need to meet the following requirements: (1)~some vertices may be inactive in a superstep and our solution should not generate outgoing messages for them; (2)~some Pregel algorithms perform topology mutation and our solution should correctly recover $\Gamma(v)$ of every vertex $v$ from the checkpointed data; (3)~our solution should handle Pregel algorithms where a vertex $v$ needs to examine every received message in order to generate and send new messages; (4)~our solution should keep the familiar vertex-centric programming model, with only minimal additional issues that a programmer needs to take care of. As we shall see in Section~\ref{sec:lwcp}, our solution satisfies all the four requirements listed above.

Lightweight checkpointing significantly improves the checkpointing time (and hence the failure-free performance). For example, when computing PageRank on WebUK (a web graph with 5.5 billion edges), it takes around 60 seconds to write a conventional checkpoint while lightweight checkpointing takes only 2 seconds.

\vspace{1.5mm}

\noindent{\bf Log-Based Recovery.} Let us use PageRank computation as an example again. Even if only one machine crashes at superstep~17, all machine will load the checkpoint at superstep~10 to roll the states of its vertices back. Then, the recovery simply reruns from superstep~11 to superstep~17. However, the states of the vertices in the surviving machines are already at superstep~17 when the failure occurs, and hence the computation of recovering the states of these vertices is redundant.

To avoid the above problem, \cite{ftgiraph} proposed a message-log based method for faster recovery, which does not roll the states of surviving vertices back, and only reassigns the vertices in the crashed machine to another healthy machine (called a replacing machine) and reruns their computation. However, when a vertex $v$ in the replacing machine reruns its computation at a superstep, say, superstep~12, $v$ also needs to receive messages from surviving vertices. For this goal, \cite{ftgiraph} proposed to log the messages that every vertex sends at every superstep to the local disk. When recovering superstep~12, a surviving vertex can now simply load from the local disk those messages that it sent to the crashed machine at superstep~12, and re-send them to the replacing machine. The recovery is much faster since the communication involves only those messages that are sent to the replacing machine. Meanwhile, \cite{ftgiraph} observed that for a common cluster connected with Gigabit Ethernet, writing messages to local disks is much faster than sending messages over the network, and hence message logging incurs negligible overhead during the failure-free execution.

However, we find that message logging does not slow down failure-free performance only if garbage collection is not considered, but without garbage collection the disk space can be used up quickly during computation. Moreover, when the job ends, the logged messages need to be garbage collected sooner or later, which is still time consuming. To see why garbage collection is necessary, let us consider PageRank computation again. In each superstep, a message is sent along every edge, and thus the size of the logged messages is comparable to the graph size. If the computation runs for 100 supersteps before convergence, then the logged data have a size of about 100 times that of the graph itself. In contrast, if we delete all logged messages right after a checkpoint is written, then only messages logged after the latest checkpoint have to be kept for recovery, whose amount does not exceed 10 times the graph size. Unfortunately, deleting the messages logged for the previous 10 supersteps is also time-consuming, since the OS needs to traverse all data-block pointers in the related files. As a result, the garbage collection cost of message-log based recovery during the failure-free execution outweighs the benefit of faster recovery when a failure does happen (which is infrequent).

Our solution, vertex-state logging, solves the problem by only logging vertex states to local disks. When a surviving vertex needs to send messages to a replacing machine, the messages are re-generated from the logged vertex states. Since the data volume of vertex states is much smaller than that of messages, garbage collecting them is much faster and incurs negligible overhead during the failure-free execution. To our knowledge, vertex-state logging is the only approach that achieves faster recovery without sacrificing the more important failure-free performance.

\vspace{1.5mm}

\noindent{\bf Contributions and Paper Organization.} While existing Pregel-like systems simply implement the straightforward heavyweight checkpointing method~\cite{pregel}, our work explores the properties specific to the framework of Pregel to significantly improve the performance of checkpointing and recovery. Specifically, we eliminate the need of storing messages in both a checkpoint and local logs, by generating messages online from vertex states; we also significantly reduce the amount of edge data stored in checkpoints and local logs, using the idea of incremental checkpointing~\cite{survey}. Robust solutions are developed to make the ideas work for a generic Pregel algorithm, while keeping the user-friendly vertex-centric programming interface with minimal additional issues that users need to take care of. Our vertex-state logging approach also solves the problem of expensive garbage collection cost suffered by a recently proposed log-based recovery approach, making log-based recovery truly beneficial in practice.

Our solutions were implemented on top of an efficient open-source Pregel system called Pregel+~\cite{pregelplus}, which has also been used in many recent works~\cite{blogel,exp,ppa}. The implementation utilizes the latest technologies such as the new ULFM (User-Level Failure Mitigation) standard of MPI for both efficiency and portability. Extensive experiments were conducted to verify the efficiency of our solutions. All system and application codes are open source and can be downloaded from \url{http://www.cse.cuhk.edu.hk/pregelplus/ft.html}.

The rest of this paper is organized as follows. We review the related work in Section~\ref{sec:related}. Our basic fault-tolerant framework is presented in Section~\ref{sec:basic}. We then introduce our solution to lightweight checkpointing in Section~\ref{sec:lwcp}, and the extension to support log-based recovery in Section~\ref{sec:log}. Experimental results are reported in Section~\ref{sec:results} and we conclude the paper in Section~\ref{sec:conclude}.

\section{Related Work}\label{sec:related}
We first review the framework of Pregel and Pregel-like systems. Then, we discuss related work on fault tolerance for general distributed systems and for Pregel-like systems.

In this paper, we assume that the input graph $G=(V, E)$ is stored on HDFS, where each vertex $v\in V$ has a unique ID $id(v)$ (we use $v$ and $id(v)$ interchangeably for simplicity) and an adjacency list $\Gamma(v)$. If $G$ is undirected (resp.\ directed), $\Gamma(v)$ contains all $v$'s neighbors (resp.\ out-neighbors). In Pregel, each vertex $v$ also maintains (1)~a value $a(v)$ which gets updated during computation, and (2)~a label $active(v)$ indicating whether $v$ is active or halted in the current superstep. Let us define the state of $v$ in Pregel as a triple $state(v)=(a(v), \Gamma(v), active(v))$. A Pregel program is run on a cluster of worker machines (or simply workers), denoted by $\mathbb{W}$.

\subsection{Pregel \& Pregel-Like Systems}\label{ssec_pregel}
\noindent{\bf Computation Model of Pregel.} A Pregel program starts by loading the input graph $G$ from HDFS, where each vertex $v$ is distributed to a worker $W\in\mathbb{W}$ according to a partitioning function $hash(.)$. Specifically, a vertex $v$,  along with its adjacency list $\Gamma(v)$,  is assigned to worker $W=hash(v)$. We denote the set of all vertices that are assigned to worker $W$ by $V_W$.

In Pregel, a user needs to specify the behavior of a vertex $v$ in a user-defined function (UDF) {\em compute}({\em msgs}), where {\em msgs} is the set of messages received by $v$, which were sent from other vertices in the previous superstep. In $v$.{\em compute}(.), $v$ may update $a(v)$ and $\Gamma(v)$, send messages to other vertices, and vote to halt (i.e., deactivate itself). Only active vertices will call {\em compute}(.) in a superstep, but a halted vertex will be reactivated if it receives a message. The program terminates when all vertices are halted and there is no pending message for the next superstep, and then the results (e.g. $a(v)$ of every vertex $v$) are dumped to HDFS.

Conceptually, the computation logic of $v$.{\em compute}(.) can be formulated as the following function:
\begin{equation}\label{eq:f}
(state^{(i)}(v), M^{(i)}_{out}(v))\gets f(id(v), state^{(i-1)}(v), M^{(i)}_{in}(v)),
\end{equation}
where we use superscript~$(i)$ to indicate the corresponding superstep number (i.e., $i$). Specifically, (1)~$state^{(i)}(v)=(a^{(i)}(v), \Gamma^{(i)}(v), active^{(i)}(v))$ refers to the state of $v$ after its computation at superstep~$i$; (2)~$M^{(i)}_{in}(v)$ refers to the set of messages received by $v$ at the beginning of superstep~$i$; and (3)~$M^{(i)}_{out}(v)$ refers to the set of messages sent by $v$ in superstep~$i$.

Since Pregel adopts a synchronous execution model, after $v$.{\em compute}(.) is called on every active vertex $v$, the outgoing messages (i.e., $M^{(i)}_{out}(v)$ of every $v$) need to be completely shuffled from the sender side to the receiver side (i.e., $M^{(i+1)}_{in}(u)$ of every $u$) before the next superstep begins.

Users may also implement a message combiner to specify how to combine messages that are sent to the same vertex $u$, so that on a worker $W$, the outgoing messages to be sent by vertices in $V_W$ to $u$ will be combined into a single message by $W$ locally, and then sent to $u$. For example, in PageRank computation, the combiner can be implemented as the summation operation, since only the sum of incoming messages is of interest in {\em compute}(.). Message combiner effectively reduces the number of messages transmitted.

Pregel also allows users to implement an aggregator for global communication. Each vertex can provide a value to an aggregator in {\em compute}(.) in a superstep. The system aggregates those values and makes the aggregated result available to all vertices in the next superstep.

\vspace{1.5mm}

\noindent{\bf Pregel-Like Systems.} Many Pregel-like systems have been developed in recent years, which adopt the user-friendly vertex-centric model of Pregel. Some systems follow a similar design as Pregel, such as Giraph~\cite{giraph}, GPS~\cite{SalihogluW13ssdbm} and Pregel+~\cite{pregelplus}, and perform synchronous execution, with vertices communicating with each other by message passing. There are also vertex-centric systems that follow a different design from Pregel, such as GraphLab~\cite{graphlab} and its subsequent version PowerGraph~\cite{powergraph} (both systems are simply called GraphLab). GraphLab adopts a shared memory abstraction where a vertex $v$ directly accesses the data of its adjacent vertices and edges (or their replicas on $v$'s machine). GraphLab also schedules vertices for processing in an asynchronous manner, which leads to faster convergence for some algorithms where vertex
values converge asymmetrically. However, \cite{tamerExp} and~\cite{exp} discover that GraphLab's asynchronous mode is generally slower than its synchronous mode that simulates the framework of Pregel, due to the overhead of enforcing data consistency under race conditions (e.g., by using locks).

This paper mainly focuses on fault tolerance issues under the synchronous computation model of Pregel, but the idea of generating messages from vertex states can be easily extended to work under the asynchronous model of GraphLab. In the remainder of this section, we discuss related work on fault tolerance, while we refer interested readers to\cite{tamerExp} and~\cite{exp} for more detailed reviews on Pregel-like systems.

\subsection{Related Work on Fault Tolerance}\label{ssec:ft}
The studies of fault tolerance in distributed message-passing systems date back to the 80s--90s, and the techniques are well surveyed in~\cite{survey}, including coordinated checkpointing, uncoordinated checkpointing and incremental checkpointing. Existing Pregel-like systems adopt coordinated checkpointing, which writes a checkpoint right after a synchronization barrier (i.e., end of message shuffling). Uncoordinated checkpointing is more efficient for asynchronous computation models, one representative algorithm of which is the Chandy-Lamport snapshot~\cite{distsnap}. For example, GraphLab adapts the Chandy-Lamport snapshot to incrementally construct each consistent snapshot without suspending execution~\cite{graphlab}. Incremental checkpointing reduces the amount of data in a checkpoint, by avoiding rewriting portions of states that do not change between consecutive checkpoints. For example, if the topology of a graph does not change throughout the computation, there is no need to write edges to any checkpoint other than the first one. However, existing Pregel-like systems have not even considered this simple version of incremental checkpointing.

Chandy-Lamport snapshot~\cite{distsnap} can be used for checkpointing asynchronous vertex-centric computation like that of GraphLab. In this approach, a checkpointing request is initiated at fixed intervals, where each worker schedules the saving of the current states of its vertices to HDFS one by one. However, the saved states may be inconsistent. To see this, consider two vertices $u$ and $v$, and assume that the following four events happen in order: (1)~$u$'s state is saved, (2)~$u$ updates $a(u)$ and sends a message to $v$, (3)~$v$ receives the message and updates $a(v)$, (4)~$v$'s state is saved. Then, any snapshot containing the saved states of $u$ and $v$ is inconsistent, since $a(u)$ refers to the old value before Event~(2), but $a(v)$ is affected by the updated value of $a(u)$ after Event~(2). To prevent the above inconsistency, whenever a vertex $u$ is checkpointed, it broadcasts a checkpointing request to all its neighbors, before sending any messages. When a vertex $v$ receives a checkpointing request, it ignores the request if it is already checkpointed; otherwise, $v$ saves its state and broadcasts a checkpointing request to all its neighbors. In the previous example, $u$ will now send a checkpointing request to $v$ before Event~(2), and thus $v$ will save its state before Event~(3) (assuming communication channels are FIFO). As a result, both $a(u)$ and $a(v)$ do not reflect the effect of Event~(2) and are thus consistent. We remark that our idea of generating messages from vertex states is also applicable to Chandy-Lamport snapshot, where when a vertex $v$ saves its state, it does not need to save the incoming messages since outgoing messages can be generated without them.

The message logging method of~\cite{ftgiraph} has been described in Section~\ref{sec:intro}, where we assume that a checkpoint is written every 10 supersteps, and a failure occurs at superstep~17. However, the algorithm becomes more complicated if cascading failures are considered, as the states of the vertices may be at more than 2 different supersteps. For example, assume that the first failure happens on worker $W_1$, and then during recovery, another failure happens at superstep~15 on $W_2$. In this case, (1)~the states of vertices on $W_1$ (which replaces the crashed worker at the first failure) are at superstep~15; (2)~vertices on $W_2$ are reassigned to another machine that loads their states at superstep~10 from the latest checkpoint on HDFS; (3)~the states of all other vertices are at superstep~17. To be robust to cascading failures, \cite{ftgiraph} classifies vertices by their states, and requires a vertex whose state is at superstep~$i$ to perform vertex-centric computation only after superstep~$i$ is recovered. However, the algorithm of \cite{ftgiraph} only considers the logic related to message passing, while we consider a more complete solution to the framework of Pregel in general, including other aspects such as the recovery of aggregator, and garbage collection.

We note that \cite{ftgiraph} emphasizes more on vertex reassignment strategies, and does not discuss important issues such as the recovery of aggregator and garbage collection, which we cover in this paper as a complete solution of log-based recovery. When a failure happens, \cite{ftgiraph} reassigns vertices in crashed workers to the surviving workers using a cost-sensitive reassignment algorithm, to achieve parallelism in recovery. The reassignment is computed by the master and written to a zookeeper; each worker then obtains the reassignment from the zookeeper and loads the assigned vertices. This solution for reassignment, however, changes the vertex partitioning function $hash(.)$, which is often a simple hash function on vertex ID. In contrast, our recovery solution endeavors to retain the $hash(.)$ function.

Among other related work, the fault-tolerance protocols surveyed in~\cite{survey} are mainly designed for a general message-passing system, and are transparent to the concrete computation model. As a result, compared with tailor-made solutions to the framework of Pregel, the general-purpose protocols incur additional overhead like piggybacked information and dependency tracking. At the other extreme, efficient recovery methods have been designed to eliminate the need of checkpointing, but they either have less expressiveness, or incur much additional burden during computation. For example, if an algorithm is self-correcting towards a fix point, optimistic recovery~\cite{optimistic} simply re-initiates the states of vertices in crashed workers and continues execution. Another work, \cite{replication}, avoids checkpointing by constructing $k$ replicas for each vertex on different workers, and relies on replicas for recovery. This solution is resilient to the failure of $(k-1)$ machines. However, replicas consume additional memory space, and the approach still slows down the failure-free performance since any update to a vertex needs to be synchronized to all its replicas.


\section{The Framework for Fault Recovery}\label{sec:basic}

This section presents our fault-tolerant framework on top of which we implement our fault recovery algorithms. The framework is implemented upon the Pregel+ system~\cite{pregelplus} (note that our framework is general and can be applied to any Pregel-like systems). Pregel+ implements the communication operations using Message Passing Interface (MPI). Implementing the communication layer by MPI provides advantages such as high efficiency and portability. Moreover, MPI programmers do not need to take care of the mapping between each computing process to the concrete machine. All that is necessary is to specify the number of processes to run, and the hostnames of the machines that the job is to run on. The concrete MPI implementation (e.g., OpenMPI, MPICH) will assign the processes to the machines (e.g., in a round-robin fashion) and start them automatically. The processes are numbered by 0, 1, 2, $\cdots$, where the ID of a process is also called its rank. The rank-to-machine mapping is automatically tracked by MPI and is transparent to programmers. This feature increases the portability of our fault-tolerant framework, allowing it to run on any number of machines deployed with any platform (with MPI installed) without any additional effort.

We now introduce the design of our fault-tolerant framework.

\vspace{1.5mm}

\noindent{\bf Worker Reassignment.} As we mentioned in Section~\ref{ssec:ft}, our framework is designed to retain the same vertex partitioning function $hash(.)$ even after recovery. Meanwhile, even if there is no standby machine, we do not want to overburden a surviving machine by assigning all vertices in a crashed machine to it. For this purpose, we partition the input graph into $n$ parts, where $n$ is a multiple (e.g., $c$ times) of the number of machines, and each part is assigned to one process (or worker). As a result, each machine runs $c$ workers, and each worker is responsible for every vertex $v$ such that $hash(v)$ equals the rank of the worker. If a machine is down, the $c$ workers can be reassigned to as many as $c$ different machines, so that a surviving worker will be assigned only around $1/c$ more workload than before. Moreover, $hash(.)$ remains the same since the rank-to-machine mapping is automatically tracked by MPI.

We remark that $hash(.)$ is a frequently evaluated function in Pregel, and it is important to keep the function simple (and hence complicated reassignment strategy should be avoided). This is because, when a vertex needs to send a message to another vertex $v$, it needs to compute the worker ID $W=hash(v)$ and then send the message to $W$. In practice, the message is appended to a message queue, which buffers all messages that are to be sent to $W$. Each worker of Pregel+ maintains $|\mathbb{W}|$ outgoing message queues, one for each worker in the worker set $\mathbb{W}$. When all active vertices have been processed for a superstep, the generated messages in each queue are then combined and sent to the target worker in one batch.

\vspace{1.5mm}

\noindent{\bf Commits.} Although messages in a queue can be sent to the target worker (in smaller batches) in parallel with the vertex-centric computation so that the network bandwidth is also utilized during the computation, the effect of message combining is reduced since messages in different batches cannot be combined. Moreover, the vertex-centric computation is often lightweight, i.e., the cost of generating a message is negligible compared with the cost of transmitting the message. Therefore, in each superstep, we adopt the simple workflow of computation followed by communication: (1)~vertex-centric computation is performed first to generate all out-going messages, and then (2)~the generated messages are combined and sent to target machines, and finally, (3)~all workers synchronize their partially aggregated data and control information, to obtain the final aggregator value and to decide whether to continue the next superstep.

Note that a worker can only detect a failure when it communicates. As a result, by performing computation before communication, it is guaranteed that when a worker $W$ detects a failure in a superstep~$i$, all vertex states and partially aggregated data and control information of $W$ have been fully updated by superstep~$i$. In this case, we say that the state of $W$, denoted by $s(W)$, is at superstep~$i$, or simply, $s(W)=i$. We also say that $W$ {\bf partially commits} superstep~$i$. Note that when a failure occurs at superstep~$i$, every worker must have partially committed superstep~$i$, and this property is important for log-based recovery which does not roll the state of a surviving worker back.

Partial commit only refers to the situation where all workers finish their computation in a superstep. If all workers also finish their communication for a superstep, which means that all messages reach the receiver side and the global aggregator value and control information are obtained, then we say that the superstep is {\bf fully committed}. We can only checkpoint a superstep~$i$ or start a new superstep (i.e., $(i+1)$), after superstep~$i$ is fully committed. This is because for every vertex $v$, $M_{in}^{i+1}(v)$ is the input to $v.${\em compute}(.) for superstep~$(i+1)$, and should be included in the checkpointed data for superstep~$i$.

\vspace{1.5mm}

\noindent{\bf Failure Detection and Error Handling.} Efficient failure detection mechanism is always an important issue of any fault-tolerant distributed system. Implementing failure detection logic (e.g., heartbeat signals) requires users to manage worker-to-machine mapping and to hardcode details like port numbers, which ruins the simplicity and portability provided by MPI. Moreover, earlier MPI libraries do not provide a mechanism to exclude the set of failed workers, and a surviving worker may at best report the detected failure and then abort. This problem hinders applications that require fault tolerance from leveraging existing MPI libraries in their implementations, forcing them to rebuild the systems from scratch and to reinvent the wheel of efficient communication primitives whose algorithms have been studied for decades and implemented in MPI libraries.

In 2012, a working group of the MPI Forum proposed User-Level Failure Mitigation (ULFM)~\cite{ulfm} for the MPI-3 Standard, which provides a resilience extension to MPI by including additional communication primitives with new semantics (e.g., for failure notification). ULFM has already been supported by main-stream MPI libraries such as OpenMPI\footnote{\url{http://fault-tolerance.org/}} and MPICH\footnote{\url{http://www.mpich.org/static/docs/v3.2/}}. These extended communication primitives start with prefix ``MPIX\_'' rather than the ``MPI\_'' prefix for standard primitives. Our fault-tolerant framework extends Pregel+ with the following two ULFM primitives, in order to enjoy the high portability of MPI and to leave low-level details such as worker-to-machine mapping to MPI:

\begin{itemize}
\item {\bf MPIX\_Comm\_revoke(.).} The function takes a worker set $\mathbb{W}$ as input, and is called by a worker $W\in\mathbb{W}$ to asynchronously notify every other worker in $\mathbb{W}$ about an error. Upon receiving the notification, a worker will immediately abort its on-going MPI communication primitive, and report an error. We denote the function by {\em mpi\_revoke}($\mathbb{W}$) for simplicity.
\item {\bf MPIX\_Comm\_shrink(.).} This is a collective function called by every surviving worker in a worker set $\mathbb{W}$ that detects a failure, and returns a new worker set containing all the surviving workers. We denote the function by {\em mpi\_shrink}($\mathbb{W}$) for simplicity.
\end{itemize}

Each worker calling {\em mpi\_shrink}($\mathbb{W}$) reports its own status, and if it detected the failure of another worker $W_f$ when it was communicating with $W_f$, the status of $W_f$ is also reported. The primitive {\em mpi\_shrink}($\mathbb{W}$) blocks until information about all workers in $\mathbb{W}$ are received. Notably, {\em mpi\_shrink}($\mathbb{W}$) ignores any notification asynchronously sent from any worker $W\in\mathbb{W}$, which is the key property which we use for failure notification in our framework. We shall discuss how we use these two ULFM primitives soon when we discuss the execution flow of our framework.

\vspace{1.5mm}

\begin{figure*}[t]
    \centering
    \includegraphics[width=1.48\columnwidth]{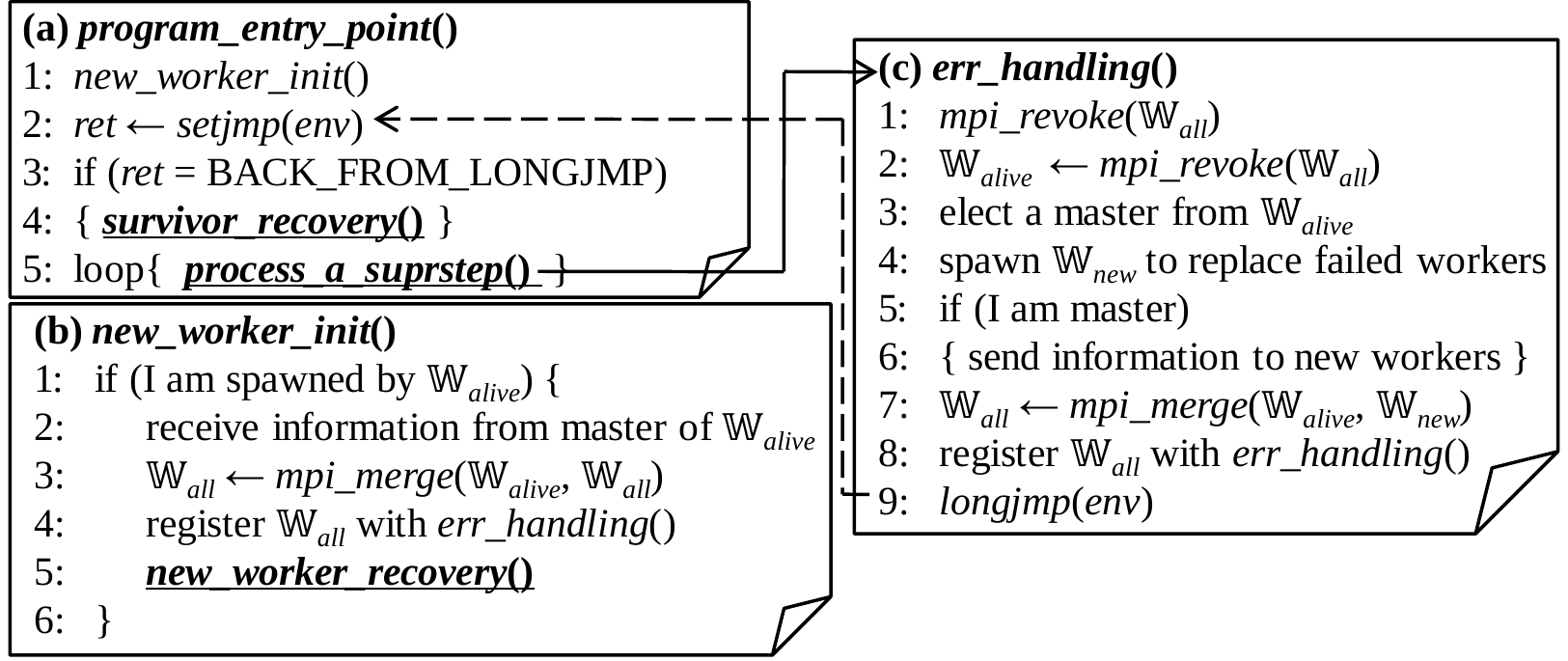}
    \caption{Framework for Fault Recovery}\label{frame}
\end{figure*}

\noindent{\bf Avoiding Single-Point-of-Failure.} An existing Pregel-like system usually runs a master, and a group of slaves that perform the actual computation. The master is responsible for monitoring the computing process to detect errors, and for aggregating partially aggregator values and control information. Master is a single point of failure (SPOF): the whole job fails if the machine running the master is crashed. Although SPOF can be mitigated by maintaining a secondary master, we adopt a more robust solution that allows any worker to be elected as a master, so that the job will not fail even if $(|\mathbb{W}|-1)$ workers fail.

We define the master as the worker $W$ with the largest state $s(W)$, i.e., the longest-living worker, with ties broken by worker ID. When we obtain the set of workers surviving a failure (using {\em mpi\_shrink}(.)), our framework will let the surviving workers immediately synchronize their states to elect a new master. The benefit of letting the longest-living worker be the master (let it be worker $W$) is that, $W$ can log the globally synchronized aggregator values and control information from $W$ during its execution, and all workers can directly obtain these global information during recovery before reaching superstep~$s(W)$. This design simplifies log-based recovery, where some workers do not perform computation and thus cannot obtain partially aggregated value and control information for synchronization, especially for the complicated case of cascading failures.

\vspace{1.5mm}

\noindent{\bf The Framework.} We now present our fault-tolerant framework, upon which we implement our fault recovery algorithms to be presented in the next two sections. Our framework deals with three worker sets: (1)~the set of all workers, denoted by $\mathbb{W}_{all}$; and when a failure occurs, (2)~the set of workers that survive the failure, denoted by $\mathbb{W}_{alive}$; and (3)~the set of new workers, denoted by $\mathbb{W}_{new}$, spawned by the surviving workers to replace the failed workers.

Before describing the execution flow of our framework, we first describe some important MPI primitives that we use. Firstly, every worker in a worker set (e.g., $\mathbb{W}_{alive}$) can call a collective function {\bf MPI\_Comm\_spawn}(.), to spawn a set of new workers (e.g., $\mathbb{W}_{new}$). The function takes arguments like the number of new workers to spawn, and a list of machines to spawn them on (e.g., in a round-robin manner). Secondly, a worker may obtain the set of workers that collectively spawn it, by calling a function {\bf MPI\_Comm\_get\_parent}(.), which returns {\em NULL} if the worker starts normally. Thirdly, worker sets (e.g., $\mathbb{W}_{alive}$ and $\mathbb{W}_{new}$) can be merged into a new worker set, by calling a function {\bf MPI\_Intercomm\_merge}(.). Finally, one can register an {\bf error handling function} to a worker set $\mathbb{W}$, so that if an on-going communication primitive over $\mathbb{W}$ reports an error, the execution flow enters the error handling function.

In our framework, every worker runs the same program (over disjoint sets of vertices) as shown in Figure~\ref{frame}. We mainly focus on the execution flow in Figure~\ref{frame}, and different recovery strategies can be implemented under this framework, by specifying different operations for the three underlined functions: (1)~{\em process\_a\_superstep}(), which specifies the algorithm for processing a superstep; and when a failure occurs and after $\mathbb{W}_{all}$ has been recovered as $\mathbb{W}_{alive}\cup\mathbb{W}_{new}$, (2)~{\em survivor\_recovery}() specifies how a surviving worker in $\mathbb{W}_{alive}$ should react to the failure, and (3)~{\em new\_worker\_recovery}() specifies how a newly-respawned replacing worker should react to restore the pre-failure state of a failed worker.

\vspace{1.5mm}

\noindent{\em \underline{Main Execution Flow}.} A worker starts by entering the main execution flow of Figure~\ref{frame}(a), where we omit details like initializing superstep number and registering an error handling function to $\mathbb{W}_{all}$ (which is {\em err\_handling}() as detailed in Figure~\ref{frame}(c)). Line~1 refers to the recovery process detailed in Figure~\ref{frame}(b), and is only run by a respawned worker. A worker that starts normally directly goes to Line~2, where it backs up the execution environment before the iterative computation in Line~5. Here, we use the {\em setjmp} and {\em longjmp} functions of the C library. If a worker calls {\em setjmp}($env$) to back up its environment to $env$, it can later call {\em longjmp}($env$) to return to the backup position. Line~3 checks whether the worker is a survivor of a failure, and just jumped back from error handling. If so, it enters Line~4 to recover its data. Finally, Line~5 performs the iterative computation, and this is where communication error may occur, in which case the execution flow will enter {\em err\_handling}().

\vspace{1.5mm}

\noindent{\em \underline{Error Handling}.} We now consider the execution flow of a worker that survives a failure. Suppose a worker $W_f$ fails, then any worker $W_d$ communicating with $W_f$ will detect the failure and call {\em err\_handling}(). In Figure~\ref{frame}(c), $W_d$ will then call {\em mpi\_revoke}($\mathbb{W}_{all}$) at Line~1 to notify other workers in $\mathbb{W}_{all}$ about the failure, and blocks on {\em mpi\_shrink}($\mathbb{W}_{all}$) at Line~2. Upon receiving the notification, a worker aborts its on-going communication, enters {\em err\_handling}() and blocks on {\em mpi\_shrink}($\mathbb{W}_{all}$). Recall that {\em mpi\_shrink}($\mathbb{W}_{all}$) ignores any revoking notification, and when all surviving workers reach {\em mpi\_shrink}($\mathbb{W}_{all}$) at Line~2, the function returns $\mathbb{W}_{alive}$ to the workers. This is because all statuses of workers in $\mathbb{W}_{all}$ have been collected, where the status of a failed worker is reported by a surviving worker that detects it.

Then, the surviving workers elect a master at Line~3, and spawn a set of $(|\mathbb{W}_{all}|-|\mathbb{W}_{alive}|)$ new workers, $\mathbb{W}_{new}$, to replace the failed ones (Line~4). The elected master then sends information to each new worker, such as the assigned worker ID and the latest checkpoint to load (Lines~5--6). Finally, a surviving worker merges $\mathbb{W}_{alive}$ and $\mathbb{W}_{new}$ as the new $\mathbb{W}_{all}$ (Line~7), and registers {\em err\_handling}() to it (Line~8). In the end, {\em longjmp} is called at Line~9 to jump back to to Line~2 of Figure~\ref{frame}(a). After jumping back, Line~4 of Figure~\ref{frame}(a) will be called where the surviving worker recovers its data (e.g., by loading a checkpoint).

\vspace{1.5mm}

\noindent{\em \underline{Execution Flow of a Respawned Worker}.} When $\mathbb{W}_{new}$ is created by Line~4 of Figure~\ref{frame}(c), we have $\mathbb{W}_{all}=\mathbb{W}_{new}$ for every respawned worker. A respawned worker enters Line~1 of Figure~\ref{frame}(a) to initialize its state, which is detailed in Figure~\ref{frame}(b). Specifically, the worker first obtains information like its assigned worker ID and the latest checkpoint (Line~2), and then incorporates $\mathbb{W}_{alive}$ into $\mathbb{W}_{all}$ (Line~3) and registers {\em err\_handling}() to it (Line~4). Finally, the worker restores the pre-failure state of a failed worker in Line~5 (e.g., by loading a checkpoint), before returning to the main execution flow for iterative computation.

\section{Lightweight Checkpointing}\label{sec:lwcp}
In this section, we describe our checkpoint-based recovery algorithms on top of our fault-tolerant framework.

\vspace{1.5mm}

\noindent{\bf Checkpointing during Normal Execution.} In the {\em process\_a\_superstep}() procedure of Figure~\ref{frame}(a), a worker $W$ processes a superstep~$i$ as follows: (1)~{\em compute}(.) is called on every active vertices in $V_W$; (2)~messages are shuffled to the receiver side, synchronization is performed to obtain the global aggregator value and control information; (3)~if the current superstep needs to be checkpointed, write the data of vertices in $V_W$ to HDFS, and then delete the previous checkpoint on HDFS. A barrier is needed before Step~(3), to guarantee that all workers have globally committed the superstep before starting checkpointing. A barrier is also needed after the current checkpoint is written and before starting to delete the previous checkpoint, to guarantee that all data of the current checkpoint is written (otherwise, the previous checkpoint is still valid), and we say that the checkpoint is committed in this case.

The condition for checkpointing is user-defined. For example, a checkpoint can be written for every $\delta$ supersteps, or every $\delta$ minutes. In the latter case, when the master fully commits a superstep, it checks whether the current time is more than $\delta$ minutes from the time of committing the last checkpoint, and if so, it notifies all workers in $\mathbb{W}_{all}$ to write a checkpoint. The time-interval based strategy is suitable for Pregel algorithms where the time taken by different supersteps varies considerably.

We denote the checkpoint for superstep~$i$ by $CP[i]$, which consists of a file on HDFS for each worker $W\in\mathbb{W}_{all}$. Specifically, each worker $W$ contributes to $CP[i]$ by writing the data of vertices in $V_W$ (after superstep~$i$ is committed) to a file denoted by $CP_W[i]$. The benefit is that when $W$ needs to roll back to superstep~$i$ later, it may simply load the file $CP_W[i]$ from HDFS.

However, at the beginning of a job, the vertices of each worker $W$ (i.e., $V_W$) may not be stored continuously in the input graph, and when different workers load different portions of the input graph, they need to shuffle the vertices with each other to obtain their own vertices for processing. To avoid the shuffling during recovery, when each worker $W$ obtains $V_W$ and before it starts iterative computation, it will write its vertex data to a file $CP_W[0]$ on HDFS as part of the initial checkpoint $CP[0]$. The iterative computation starts from superstep~1, and if a failure occurs before another checkpoint is written, each worker $W$ simply loads $CP_W[0]$ and rolls back to the beginning of the job.

\vspace{1.5mm}

\noindent{\bf Incremental Checkpointing of Edges.} A conventional checkpoint simply stores $\Gamma(v)$ of every vertex $v$, which contributes a data volume of $O(|E|)$ to the checkpoint. If $k$ checkpoints are written in a job, then $O(k|E|)$ amount of edge data are written to HDFS, which accounts for a significant portion of the failure-free execution time.

The high cost of this na\"{i}ve solution can be avoided by the idea of incremental checkpointing. For example, in PageRank computation, the graph topology is static, and thus each worker $W$ can simply load the edges from $CP_W[0]$. Thus, there is no need to store edges in any checkpoint $CP[i]$ where $i\geq 1$.

However, there also exist Pregel algorithms that perform topology mutation. A common type of algorithms only perform edge deletions during the iterative computation, such as the $k$-core finding algorithm of~\cite{socialnet}. To be both general and space-efficient, we let each worker $W$ log its requests of topology mutation to the local disk. When $W$ writes a new checkpoint, it appends the logged requests to a log file (for $W$) on HDFS, denoted by $E_W$, and then deletes the requests from the local disk.

To recover the adjacency lists of vertices in $V_W$, $W$ simply loads the initial edge data from $CP_W[0]$ and then replays the logged mutation requests loaded from $E_W$.

To see why this approach is more efficient, consider a Pregel algorithm with only edge deletions. In this case, there are at most $|E|$ mutation requests in total. In other words, at most $O(|E|)$ edge mutation data are written to HDFS throughout the computation regardless of the number of checkpoints written, and the recovery of edge data also only loads $O(|E|)$ mutation requests to replay. We remark that this bound is loose, since in reality, the mutation requests are usually much less than $O(|E|)$ (e.g., $k$-core algorithm of~\cite{socialnet}). This approach also supports log-based recovery, since a surviving worker may simply forward edge mutation requests (loaded from its local log) to failed workers.

\vspace{1.5mm}

\noindent{\bf Message Generation from Vertex States.} A conventional checkpoint of a superstep~$i$ also needs to store all messages generated in superstep~$i$. Our baseline algorithm that implements conventional checkpointing writes a checkpoint for superstep~$i$ after it is globally committed. Since messages have been combined and shuffled to the receiver side, each worker $W$ simply saves all received messages to $CP_W[i]$. To load $CP[i]$ for recovery, each worker $W$ simply loads the incoming messages from $CP_W[i]$ for use by superstep~$(i+1)$, and there is no need to shuffle messages for superstep~$i$.

We call this baseline algorithm as {\bf HWCP} (\underline{h}eavy\underline{w}eight \underline{c}heck\underline{p}ointing), since each checkpoint stores all the edges and messages, which is heavyweight. In our framework, HWCP is implemented by specifying both recovery functions {\em survivor\_recovery}() and {\em new\_worker\_recovery}() in Figure~\ref{frame} with the same logic: each worker $W$ loads data from latest heavyweight checkpoint $CP[i]$, and sets the superstep number back to $i$.

Even HWCP only stores fully combined messages, the volume can still be very large. While the number of messages generated in a superstep is exactly $|E|$ for PageRank computation, the number can be even much larger in some Pregel algorithms. For example, the triangle finding algorithm of~\cite{socialnet} sends $\Omega(|E|^{1.5})$ messages in a superstep, as we shall discuss in the Appendix. To eliminate the need of storing messages in a checkpoint, we propose to instead generate messages from vertex states online. This algorithm is called {\bf LWCP} (\underline{l}ight\underline{w}eight \underline{c}heck\underline{p}ointing), since we only write the $O(|V|)$ vertex states (without adjacency lists) and the incremental mutation requests to the checkpoint, whose volume is typically much less than $O(|E|)$. Of course, $CP[0]$ has to store all the initial edges.

Recall that in HWCP, the logic of $v.${\em compute}(.) can be formulated into the function $f(.)$ shown in Equation~(\ref{eq:f}). In contrast, LWCP formulates the computation logic by the following two functions running in order:
\begin{eqnarray}
state^{(i)}(v) & \gets & g(id(v), state^{(i-1)}(v), M^{(i)}_{in}(v)),\label{eq:g}\\
M^{(i)}_{out}(v) & \gets & h(id(v), state^{(i)}(v)).\label{eq:h}
\end{eqnarray}

Put simply, function $g(.)$ first computes a new state for $v$ from its old state and the messages received by $v$, and then function $h(.)$ generates outgoing messages solely from the new state of $v$, without examining the incoming messages.

\vspace{1.5mm}

\noindent{\bf Model Expressiveness.} The above functions might remind you of the edge-centric Gather-Apply-Scatter (GAS) model adopted by systems like PowerGraph~\cite{powergraph} and GraphChi~\cite{graphchi}. In fact, the GAS model can be expressed using these two functions: (1)~the {\em Gather} phase obtains messages from each in-edges (i.e., $M^{(i)}_{in}(v)$) and aggregates them to update the vertex state (i.e., $state^{(i)}(v)$), and (2)~the {\em Scatter} phase computes a message for each out-edge (i.e., $M^{(i)}_{out}(v)$) from the updated state of $v$ (i.e., $state^{(i)}(v)$). Therefore, the model of LWCP is at least as expressive as the GAS model.

However, our goal is to make LWCP as expressive as the computation model of Pregel, but Equations~(\ref{eq:g}) and~(\ref{eq:h}) constitute a special case of Equation~(\ref{eq:f}). We classify Pregel algorithms into three categories, and explain how LWCP fits into each category.

The first kind of algorithm is called an {\bf always-active style} algorithm, where in each superstep, every vertex is active and sends messages whose values are computed from the vertex state. PageRank computation falls into this category, and thus our LWCP algorithm can use the original {\em compute}(.) of HWCP without any modification.

The second kind of algorithm is called a {\bf traversal style} algorithm, where a vertex only sends messages if its value is updated by the incoming messages. Examples of traversal style algorithms include the Hash-Min algorithm for computing connected components~\cite{ppa} and the algorithm for computing single-source shortest paths~\cite{pregel}. For such an algorithm, users need to slightly modify the original {\em compute}(.) of HWCP in order to use LWCP. Specifically, the vertex value $a(v)$ needs to be expanded with another boolean field indicating whether the vertex value is updated. If so, $h(.)$ generates messages according to $state^{(i)}(v)$; otherwise, no message is generated.

The above two categories are actually summarized by~\cite{catchw} and covers most Pregel algorithms. In such algorithms, the outgoing messages can always be computed from vertex state without examining incoming messages. The last kind of algorithm, however, needs to examine the incoming messages in order to generate outgoing messages, and we call such an algorithm as a {\bf request-respond style} algorithm. In such an algorithm, a requesting vertex $u$ will include its ID in its message to a responding vertex $v$, so that $v$ knows whom to send its response to.

We further classify request-respond style algorithms into two types. In the first type of algorithm, a responding vertex only needs to select and react to one requesting vertex. An example is given by the bipartite matching algorithm of~\cite{pregel}, where an unmatched vertex on one side only needs to select one vertex (that sends a matching request) from the other side to match. In this problem, the vertex value $a(v)$ needs to be expanded with another field indicating the selected vertex for matching.

In the second type of request-respond style algorithm, a responding vertex needs to send response to every requesting vertex. Even worse, a vertex $v$ may receive requests from many other vertices asking for the value of $v$, and these vertices may not be $v$'s direct neighbor. This is common for Pregel algorithms that use the pointer jumping (or path doubling) technique to bound the number of supersteps, such as the S-V algorithm of~\cite{ppa} for computing connected components, and the minimum spanning forest algorithm of~\cite{gpsopt}. In these algorithms, a vertex $v$ needs to respond to more and more vertices as the computation goes on, and we cannot include all their IDs to $a(v)$.

However, such algorithms only have a small portion of supersteps where vertices send responses, and LWCP is still applicable to the other supersteps. Let us call a superstep where vertex send requests (resp.\ responses) as a requesting (resp.\ responding) superstep, then we can see that LWCP is still applicable to a requesting superstep. Therefore, our solution is to allow users to mask out those supersteps where LWCP is inapplicable (e.g., responding superstep). Our LWCP algorithm skips the checkpointing operation in a masked superstep even if the condition for checkpointing holds, and a checkpoint will be saved for the first LWCP-applicable superstep after the masked superstep.

\vspace{1.5mm}

\noindent{\bf Programming Interface.} Our LWCP algorithm does not require users to explicitly implement two UDFs for Equations~(\ref{eq:g}) and~(\ref{eq:h}). After all, it is also impossible for a superstep where LWCP is inapplicable. Instead, we let users write the familiar {\em compute}(.) function, with some additional issues in mind, which we detail below.

Firstly, a superstep that is not LWCP-applicable should be masked. We provide two methods to mask a superstep: (1)~a vertex may mask the current superstep in {\em compute}(.), and a superstep is masked if any vertex masks it; (2)~users may implement a UDF {\em LWCPable}() called at the beginning of a superstep to determine whether to disable checkpointing.

Secondly, for an LWCP-applicable superstep, users need to include additional fields into the vertex value type according to Equation~(\ref{eq:h}), and to formulate the logic in two steps, (i)~updating vertex state using incoming messages (i.e., Equation~(\ref{eq:g})), followed by (ii)~sending messages according to the updated vertex state (i.e., Equation~(\ref{eq:h})).

In the Appendix, we illustrate how to write {\em compute}(.) of LWCP for the triangle finding algorithm of~\cite{socialnet}.

Our interface design endeavors to keep the vertex-centric programming interface of Pregel, with minor additional issues user need to take care of in order to enjoy the fast checkpointing time of LWCP. In fact, {\em compute}(.) of PageRank computation is exactly the same for both LWCP and HWCP. Although for some Pregel algorithms, a user needs to slightly modify {\em compute}(.) according to the semantics of Equations~(\ref{eq:g}) and~(\ref{eq:h}), we believe the additional workload to be reasonable just like any pay-as-you-go extensions to Pregel. For example, the recently proposed block-centric frameworks perform much better than the vertex-centric model, but they require users either to write additional computing logic inside a block~\cite{giraph++,blogel}, or to specify a scheduler that schedules vertex-centric computation inside a block~\cite{grace}. After all, if a user does not want to consider additional issues, he may simply use our HWCP algorithm.

\vspace{1.5mm}

\noindent{\bf  Algorithm of LWCP.} Similar to HWCP, our LWCP algorithm implements both recovery functions {\em survivor\_recovery}() and {\em new\_worker\_recovery}() in Figure~\ref{frame} with the same logic as follows: each worker $W$ (i)~loads the states of vertices in $V_W$ from the latest lightweight checkpoint $CP_W[i]$, (ii)~generates outgoing messages from the loaded states using the semantics of Equation~(\ref{eq:h}), and (iii)~shuffles the messages to the receiver side for use in superstep~$(i+1)$. Moreover, adjacency lists are loaded from $CP[0]$, and the logged topology mutations are replayed (if any). If there is no topology mutation, we optimize our algorithm not to load adjacency lists for surviving workers since the existing ones are valid.

Note that after loading vertex states from $CP[i]$, LWCP still needs to generate messages and then shuffle them. In contrast, HWCP directly loads the shuffled messages at the receiver side, and is thus faster. However, loading a checkpoint is just a minor and one-off cost incurred when a failure happens, and the faster checkpointing time of LWCP outweighs the slightly increased cost of checkpoint loading.

Recall that the state of a vertex $v$, $state(v)$, consists of $a(v)$, $active(v)$ and $\Gamma(v)$. We only store $a(v)$ and $active(v)$ of every vertex $v$ in a checkpoint, and $\Gamma(v)$ is handled by incremental checkpointing. However, this is insufficient for message generation, since some vertices may not call {\em compute}(.) in a superstep, because it is inactive and does not receive any message. Let us define a boolean field $comp^{(i)}(v)$ indicating whether {\em compute}(.) is called on a vertex $v$ in superstep~$i$. We store three fields into a checkpoint $CP[i]$ for each vertex $v$: (1)~$a^{(i)}(v)$, (2)~$active^{(i)}(v)$ and (3)~$comp^{(i)}(v)$. After loading them from $CP[i]$ during recovery, our LWCP algorithm generates messages for a vertex $v$ only if $comp^{(i)}(v)=$ {\em true}. Note that $active^{(i)}(v)$ cannot replace $comp^{(i)}(v)$ since a vertex $v$ may perform computation and vote to halt at last.

\vspace{1.5mm}

\noindent{\em\underline{Transparent Message Generation}.} There is yet one remaining problem: users only specify a UDF {\em compute}(.) whose semantics includes both Equations~(\ref{eq:g}) and~(\ref{eq:h}), but in Step~(ii), we only want to generate messages for a vertex using Equation~(\ref{eq:h}). Note that it is also incorrect to generate the outgoing messages of a vertex $v$ by directly calling $v.${\em compute}(.). This is because the values of $a(v)$ and $active(v)$ was loaded from the checkpoint and are thus already up-to-date, and the computing logic of Equation~(\ref{eq:g}) will change them again.

Our solution is still to generate messages for each vertex $v$ using the same UDF {\em compute}(.), but in this stage, our framework will ignore any update to the state of $v$ when users call functions like {\em set\_value}(.) and {\em vote\_to\_halt}(.) in {\em compute}(.). As a result, messages are correctly generated using the vertex states loaded from the checkpoint, without additional effort from a programmer.

\section{Log-Based Recovery}\label{sec:log}
In this section, we first describe how the log-based recovery algorithm of~\cite{ftgiraph} (that performs message logging) can be implemented under our fault-tolerant framework, and then describe a new vertex-state logging approach that further avoids expensive garbage collection during normal execution. We remark that log-based recovery algorithms also perform checkpointing, but they additionally log messages (or vertex states) to local disks. Moreover, since our vertex-state logging approach also requires users to formulate {\em compute}(.) according to Equations~(\ref{eq:g}) and~(\ref{eq:h}), it uses LWCP for checkpointing.

\vspace{1.5mm}

\noindent{\bf The Message Logging Approach.} We now present the algorithm for the message logging approach of~\cite{ftgiraph} under our framework. We denote this algorithm by {\bf HWLog}, which performs both HWCP and message logging. All current local logs are garbage collected by the respective workers after a new checkpoint is written (and committed).

For checkpointing-only algorithms, in superstep~$i$, every worker $W$ performs vertex-centric computation and updates its state $s(W)$ from $(i-1)$ to $i$. However, in log-based recovery, some workers may have $s(W)>i$ since the states of surviving workers are not rolled back, and these workers simply forward messages loaded (or generated) from local logs to those workers that perform vertex-centric computation.
Therefore, each worker needs to keep track of the states of every worker in $\mathbb{W}_{all}$ in order to decide whether to send messages to them. When $\mathbb{W}_{all}$ is recovered as $\mathbb{W}_{alive}\cup\mathbb{W}_{new}$ after a failure, the workers need to synchronize their states with each other. Here, synchronization is necessary since surviving workers can be at different supersteps due to cascading failures, and a respawned worker has to get the states of all surviving workers in order to compute the new master.

In a superstep~$i$, if a worker $W$ performs vertex-centric computation (which forwards $s(W)$ from $(i-1)$ to $i$), the generated messages to be sent to each worker $W'\in\mathbb{W}_{all}$ are buffered in a message queue and combined. The combined messages are then sent to $W'$ and meanwhile, written to a file $log_W[i][W']$ on local disk in parallel. Since local disk write is typically much faster than network transmission when Gigabit Ethernet is used, log writing usually finishes much earlier than message transmission.

We regard superstep~$i$ as partially committed by $W$ only if $log_W[i][W']$ is fully written for every $W'\in\mathbb{W}_{all}$, since the file may need to be loaded by $W$ as a whole to be forwarded to $W'$ during recovery. As a result, if a failure happens, {\em err\_handling}() needs to wait until the surviving worker finishes its asynchronous log writes before starting its error handling. A worker also needs to guarantee that the log writes are complete before fully committing a superstep, though this normally adds no overhead since message transmission is slower.

Also note that we store messages for each superstep and each destination worker in a file, so that in a recovery superstep~$i$, a survivor may simply load the messages in $log_W[i][W']$ to be sent to only those workers $W'$ that perform vertex-centric computation.

\vspace{1.5mm}

\noindent{\bf Algorithm of HWLog.} We now present our HWLog algorithm by specifying the three underlined functions in Figure~\ref{frame}. We first describe the algorithm of {\em process\_a\_superstep}(). Let the current superstep number be~$i$. There are three cases.

{\em Case 1}: $s(W)\geq i$. In this case, $W$ is a survivor who has partially committed superstep~$i$ before, and thus it does not need to perform vertex-centric computation. Instead, it loads messages from $log_W[i][W']$ for each target worker $W'$ such that $s(W')\leq i$, and sends them to $W'$. This is because such a worker $W'$ will perform computation at the next superstep (i.e., $(i+1)$), which requires these messages.

{\em Case 2}: $s(W)=i-1$. In this case, $W$ needs to perform vertex-centric computation and updates its state from $(i-1)$ to $i$. All generated messages need to be logged, since any worker may fail later and request messages from $W$ for re-computation. However, like in Case~1, only those messages for a worker $W'$ with $s(W')\leq i$ are actually sent.

{\em Case 3}: $s(W)<i-1$. This case is impossible, which can be proved by induction on $i$, using the fact that in Case~2, if the state of a worker is less than the current superstep~$i$, it will perform computation and update its state to $i$.

Finally, let $W_{mst}$ be the elected master, then if $i<s(W_{mst})$, there is no need to perform synchronization among workers to obtain the global aggregator value and control information, since they have been logged by $W_{mst}$ and thus can be directly obtained from $W_{mst}$. In contrast, when $i=s(W_{mst})$, synchronization is necessary for recovering superstep~$i$, since the earliest failure occurs in that superstep and thus $W_{mst}$ is not globally committed. Note, however, that $W_{mst}$ is guaranteed to be locally committed and should have logged the partially aggregated value and control information, which are used for the synchronization.

When a failure happens, {\em survivor\_recovery}() (in Figure~\ref{frame}(a)) retains the state of the surviving worker, but sets the superstep number back to the latest checkpointed one. The message queues are only emptied to remove on-the-fly messages, so that these queues can be used to accommodate messages read from local logs during later recovery.

In contrast, {\em new\_worker\_recovery}() (in Figure~\ref{frame}(b)) sets both the state of a respawned worker and the superstep number to the latest checkpointed superstep, and it also loads the latest checkpoint, which contains incoming messages for the next superstep as the checkpoint is heavyweight.

\vspace{1.5mm}

\noindent{\bf The Vertex-State Logging Approach.} We now present our vertex-state logging approach that improves the message logging baseline (i.e., HWLog). We call this algorithm as {\em LWLog}, which performs both LWCP and vertex-state logging. We remark that the programming interface of LWLog is exactly the same as that of LWCP described in the previous section, i.e., users write {\em compute}(.) but formulate their program with Equations~(\ref{eq:g}) and~(\ref{eq:h}).

Compared with HWLog, the content written to a local log has a much smaller data volume because messages are not included. Specifically, for each vertex $v$, only $comp^{(i)}(v)$ and $a^{(i)}(v)$ are logged. If a worker needs to generate messages of superstep~$i$ for forwarding, it generates messages for a vertex $v$ only if $comp^{(i)}(v)=$ {\em true}. Unlike LWCP, a local log does not need to store $active^{(i)}(v)$ since the logged states are just for message generation and do not overwrite the current vertex states.

To generate messages from vertex states, LWLog also uses {\em compute}(.) by temporarily ignoring updates to vertex states. There are two places that require message generation, where we denote the latest checkpointed superstep by $s_{last}$.

{\em Place 1:} when a failure occurs, a respawned worker loads $CP[s_{last}]$ and uses the loaded vertex states to generate outgoing messages for sending, which is the same as in LWCP. In contrast, error handling is triggered on a surviving worker, which directly loads messages of superstep~$s_{last}$ from its local log for sending. This is possible because LWLog adopts a slightly different garbage collection strategy from HWLog: when a new lightweight checkpoint $CP[i]$ is written, all local logs written before superstep~$i$ are deleted, but the logs written at superstep~$i$ is retained (for use by error handling later). The only exception is at the beginning of the job, where a survivor loads the initial vertex states from $CP[0]$ on HDFS rather than from any local log.

{\em Place 2:} During a recovery superstep $i$ starting from $(s_{last}+1)$, a worker that needs to forward messages simply loads the proper local vertex-state log file(s) and generates messages from the loaded vertex states for sending.

Finally, we discuss how LWLog handles a masked superstep that is not LWCP-applicable. Since the outgoing messages depend on the incoming messages in such a superstep, they cannot be recovered only from the vertex states, and therefore LWLog switches temporarily to message logging instead of vertex-state logging if a superstep is masked.

\section{Experiments}\label{sec:results}
We now report the performance of both checkpointing-based algorithms (1)~HWCP and (2)~LWCP, and log-based algorithms (3)~HWLog and (4)~LWLog. Our focus is on checkpointing time and recovery time.

All experiments were run on a cluster of 15 machines connected by Gigabit Ethernet, each with 12 cores (two Intel Xeon E5-2620 CPU) and 48GB RAM. We ran 8 workers on each machine, and thus 120 workers in total.

All our system and application code can be accessed from \url{http://www.cse.cuhk.edu.hk/pregelplus/ft.html}.

\begin{table}[t]
    \centering
	\caption{Real Graph Datasets}\label{data}
    \includegraphics[width=\columnwidth]{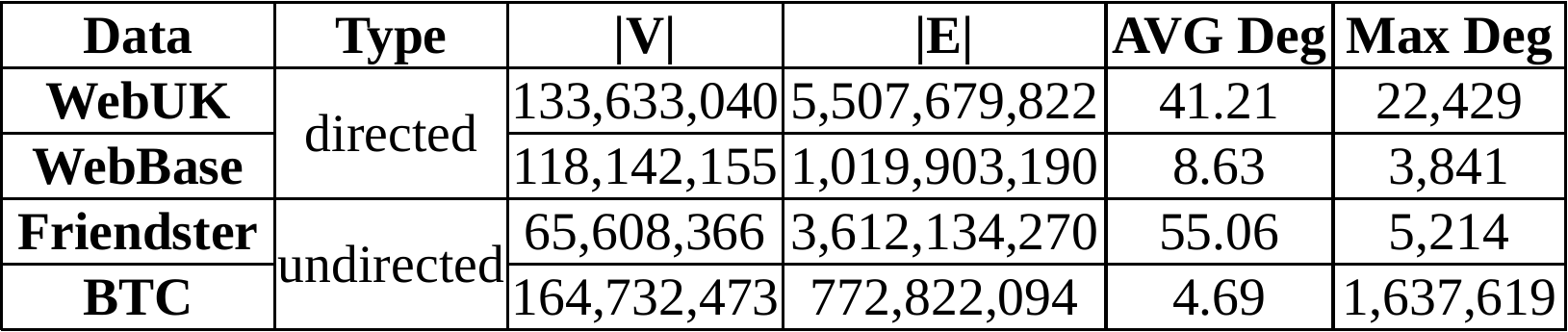}
\end{table}

\vspace{1.5mm}

\noindent{\bf Datasets.} Table~\ref{data} shows the datasets used in our experiments, including two web graphs {\em WebUK}\footnote{\url{http://law.di.unimi.it/webdata/uk-union-2006-06-2007-05}} and {\em WebBase}\footnote{\url{http://law.di.unimi.it/webdata/webbase-2001}}, one social network {\em Friendster}\footnote{\url{http://snap.stanford.edu/data/com-Friendster.html}} and one RDF graph {\em BTC}\footnote{\url{http://km.aifb.kit.edu/projects/btc-2009}}.

\vspace{1.5mm}


\noindent{\bf Algorithms.} Fault tolerance is most useful for long-running jobs. We consider two well-known long-running Pregel algorithms, PageRank computation and triangle finding. We briefly introduce them below.

PageRank computation runs considerably longer than other graph algorithms such as the computation of connected components~\cite{ppa} or single-source shortest paths~\cite{pregel}, since PageRank may take many supersteps before convergence. For a large graph, each superstep can be time consuming. Since the time of a superstep is relatively stable throughout the computation, it is common to write a checkpoint for every $\delta$ supersteps.

Triangle finding generates huge amounts of intermediate messages during the computation. For example, in the algorithm of~\cite{socialnet}, to find a triangle, $\triangle v_1v_2v_3$ (assuming $v_1<v_2<v_3$), vertex $v_1$ needs to send a message to $v_2$ asking it whether $v_3\in\Gamma(v_2)$. Since a graph can have $O(|E|^{1.5})$ triangles~\cite{triangle}, the message volume is at least $O(|E|^{1.5})$, which is superlinear to the graph size. Finding all triangles in one round leads to long-running supersteps that are susceptible to machine failures and expensive re-computation. Moreover, the aggregated memory in the cluster may not be sufficient to buffer all the messages. To solve the above problem, existing work has been considering multi-round solutions with disk-based MapReduce~\cite{triaMapred}, where each round only computes a fraction of triangles. In the Appendix, we extend the triangle finding algorithm of~\cite{socialnet} to run in multiple rounds with bounded message number in each round, whose variation for triangle counting was used in our experiments to eliminate the cost of saving enumerated triangles to disks.

Since PageRank is designed for (directed) web graphs, we ran its experiments on the two directed graphs, {\em WebUK} and {\em WebBase}. In contrast, triangle counting is normally computed in an undirected graph, and thus we ran its experiments on the two undirected graphs {\em Friendster} and {\em BTC}.

\subsection{Experiments on PageRank Computation}
In this set of experiments, we ran the PageRank algorithm of~\cite{pregel}, and wrote a checkpoint for every 10 supersteps. Since every superstep generates the same number of messages (one on each edge) during normal execution, the running time of a superstep is stable. This also holds during the recovery stage, and thus we report the average running time of a superstep for each stage of computation.

In the experiments, we killed a worker at superstep~17 to simulate a worker failure. This leads to 4 difference stages listed in order as follows, which gives us four time metrics about the running time of a superstep:
\begin{itemize}
\item {\em Stage 1:} the job first executes normally from superstep~1 to superstep~16, and we define $\bm{T_{norm}}$ as the running time of a superstep averaged over these 16 supersteps.
\item {\em Stage 2:} after the failure occurs at superstep~17, recovery of the latest checkpointed superstep (i.e., 10) is triggered. To recover superstep~10, in HWCP and LWCP, every worker loads $CP[10]$ from HDFS; while in HWLog and LWLog, only respawned workers load $CP[10]$. Moreover, since LWCP and LWLog load only vertex states, they need to generate messages and shuffle them to the receiver side. We denote the time of recovering superstep~10 by $\bm{T_{cpstep}}$, which represents the time of recovering the latest checkpointed superstep (including the time of loading the checkpoint).
\item {\em Stage 3:} after recovering superstep~10 in Stage~2, the job reruns from superstep~11 to superstep~16. We define $\bm{T_{recov}}$ as the running time of a superstep averaged over these 6 supersteps. Note that since HWLog and LWLog only transmit messages to one respawned worker that replaces the killed one, messages to the other 119 surviving workers do not need to be transmitted and thus $T_{recov}$ is expected to be much shorter than $T_{norm}$.
\item {\em Stage 4:} finally, the recovery reaches superstep~$17$, and we denote the time of recovering this superstep by $\bm{T_{last}}$. This metric represents the time of recovering the superstep where the failure occured. We separate $T_{last}$ from $T_{recov}$ since even HWLog and LWLog have to transmit all messages in this superstep. This is because after superstep~17, the job returns to normal execution starting from superstep~18, and thus every worker performs vertex-centric computation (whose performance is already captured by $T_{norm}$). We expect $T_{last}$ to be close to $T_{norm}$ since all messages are transmitted, but slightly shorter since survivors do not perform computation.
\end{itemize}

Among the metrics, $T_{norm}$ is averaged over 16 supersteps while $T_{recov}$ is averaged over 6 supersteps, which is good enough since the time of a superstep is stable in each stage. Also note that not all metrics we reported are equally important. For example, $T_{recov}$ is important in demonstrating how log-based recovery reduces the recovery time in HWLog and LWLog, while $T_{last}$ is less important as it is only related to the last superstep of recovery. Moreover, $T_{norm}$ is only reported for comparison (e.g., with $T_{recov}$) and is obviously not to be improved by any fault-tolerance mechanism. We highlight the most important metrics in red color, in all the subsequent tables that report performance results. Finally, we remark that while our algorithms support cascading failures, they will lead to more stages and thus more time metrics to report, which are too complicated for presentation purpose and are thus avoided.

\begin{table}[!t]
    \centering
	\caption{Time Metrics for Supersteps}\label{step_pr}
    \includegraphics[width=0.66\columnwidth]{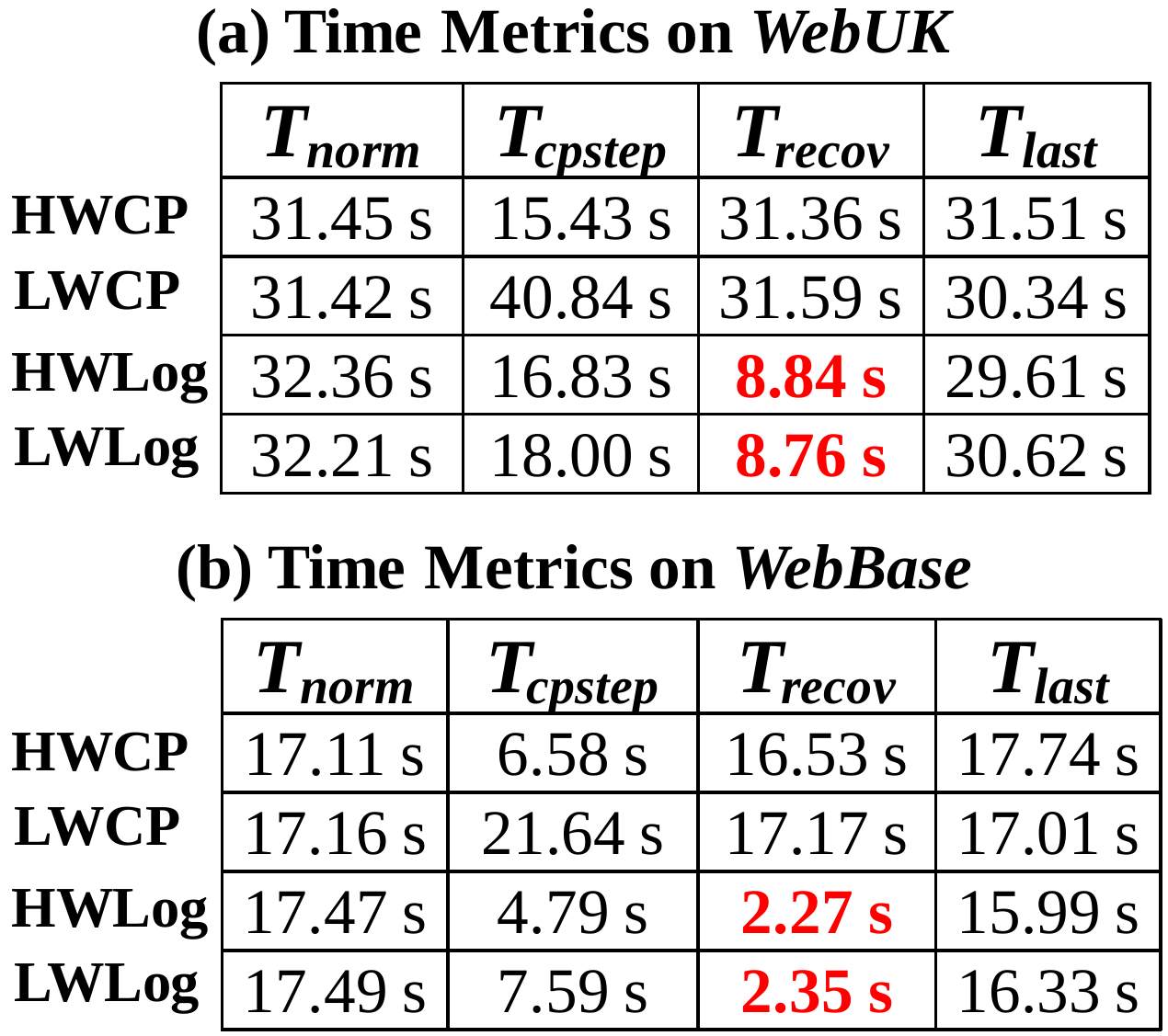}
\end{table}

\vspace{1.5mm}

\noindent{\bf Performance of Time Metrics for Supersteps.} Table~\ref{step_pr}(a) (resp.\ Table~\ref{step_pr}(b)) reports the performance of computing PageRank over {\em WebUK} (resp.\ {\em WebBase}) for the time metrics for supersteps defined above. Columns with header~$T_{norm}$ in Table~\ref{step_pr} show that during normal execution, a superstep takes around 32 seconds (resp.\ 17 seconds) on {\em WebUK} (resp.\ on {\em WebBase}).

For HWCP and LWCP, $T_{recov}$ is similar to $T_{norm}$ since checkpoint-based recovery simply reruns the computation after rolling back to superstep~$step_{CP}$. In contrast, for HWLog and LWLog, $T_{recov}$ is many times shorter than $T_{norm}$, as highlighted by the red figures in Columns with header $T_{recov}$ in Table~\ref{step_pr}. Specifically, $T_{recov}$ is around 4 times (resp.\ 8 times) shorter than $T_{norm}$ on {\em WebUK} (resp.\ {\em WebBase}). This is because log-based algorithms only transmit those messages to the respawned worker. However, recall that we only kill one of the 120 workers and thus the message volume to be transmitted is reduced to approximately 1/120 of that during normal execution. But $T_{recov}$ is not reduced to 1/120 of $T_{norm}$, which is because of two reasons: (1)~vertex-centric computation and message combining are performed in parallel during normal execution, and cannot be reduced much since the respawned worker still needs to perform these operations; (2)~only the respawned worker receives messages, which results in a communication bottleneck on the receiver side.

\begin{table}[!t]
    \centering
	\caption{Effect of Number of Failed Workers (on WebUK)}\label{trend_pr}
    \includegraphics[width=0.9\columnwidth]{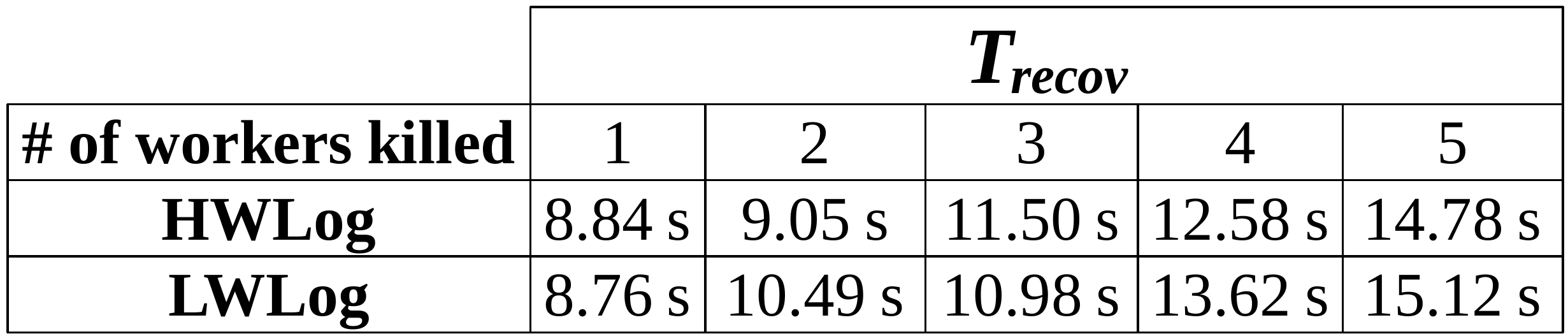}
\end{table}

Obviously, for HWLog and LWLog, if we kill more workers at superstep~17, more messages need to be transmitted during recovery and $T_{recov}$ should increase. To study the relationship between $T_{recov}$ and the number of failed workers, we repeat the previous experiments for HWLog and LWLog on {\em WebUK}, by killing more workers. The results are reported in Table~\ref{trend_pr}, where we can see that $T_{recov}$ increases slowly with the number of workers killed. In fact, $T_{recov}$ continues to increase when more workers are killed. For example, when 12 workers are killed, $T_{recov}$ is around 18 seconds for both HWLog and LWLog; while when 20 workers are killed, $T_{recov}$ is around 21 seconds for both algorithms.

Referring to Table~\ref{step_pr} again, we can see that $T_{cpstep}$ of LWCP and LWLog is longer than that of HWCP and HWLog. For example, on {\em WebUK}, $T_{cpstep}$ takes 15.43 seconds for HWCP while the time is 40.84 seconds for LWCP. This is because when recovering superstep~10 after rolling back, a worker in LWCP and LWLog needs to generate messages from vertex states (loaded from $CP[10]$ or local logs of superstep~10), and then shuffles them to the receiver side. In contrast, a worker in HWCP and HWLog directly loads incoming messages for superstep~11 from $CP[10]$.

Note that $T_{cpstep}$ is much shorter than $T_{norm}$ in HWCP and HWLog. For example, for HWCP on {\em WebUK}, $T_{cpstep}$ takes 15.43 seconds while $T_{norm}$ is 31.45 seconds. This is because incoming messages are directly loaded from $CP[10]$ when recovering superstep~10, whose time cost is much less than that of vertex-centric computation, plus that of combining and the transmission of the generated messages as required in normal execution.


In contrast, $T_{cpstep}$ is even longer than $T_{norm}$ in LWCP, since LWCP transmits the same amount of messages during recovery as in normal execution, except that these messages are generated from vertex states loaded from $CP[10]$ rather than by vertex-centric computation. Also, $T_{cpstep}$ includes the time of loading $CP[10]$ from HDFS.

However, this does not mean that LWCP is inferior to HWCP. Note that $T_{cpstep}$ is just a one-off cost for recovering a failure (which is very infrequent); while as we shall see shortly, compared with HWCP, LWCP significantly reduces the checkpointing time and thus improves the failure-free performance (of every job). Also note that the additional recovery cost incurred by $T_{cpstep}$ is limited, as $T_{cpstep}$ is close to $T_{norm}$, i.e., the time of running one superstep.

\vspace{1.5mm}

\noindent{\bf Performance of Checkpointing and Logging.} We now report the following metrics on checkpointing and logging for the same experiments described above. Note that while the previously-defined metrics mainly reflect the performance of recovery, the next few metrics reflect the additional overhead incurred by any job during failure-free execution, in order to be fault-tolerant.

\begin{itemize}
\item Since $CP[0]$ is a special heavy-weight checkpoint that includes edges but no messages, we examine the time of writing $CP[0]$, denoted by $\bm{T_{cp0}}$.
\item We also examine the time of writing a checkpoint $CP[i]$ ($i\geq 1$), which is $CP[10]$ in our experiments. We denote this time by $\bm{T_{cp}}$, which also includes the time of any garbage collection operations following the writing of $CP[10]$.
\item We examine the time of loading a checkpoint $CP[i]$ ($i\geq 1$). For our experiments, the time refers to that of loading $CP[10]$ after we kill a worker at superstep~17, which we denote by $\bm{T_{cpload}}$.

    \vspace{1mm}

    Since some workers may not load data from $CP[10]$ in log-based algorithms, and workers do not synchronize through a barrier after they load $CP[10]$, $T_{cpload}$ is averaged over the checkpoint loading time of every worker $W$ that loads data from $CP_W[10]$. Note that the time accounts for part of $T_{cpstep}$.
\item We examine the time of writing a local log, denoted by $\bm{T_{log}}$. Since some workers do not perform computation and log data, the time is averaged over all workers that write a log and over all supersteps (both in normal execution and during recovery).
\item We examine the time of loading a local log, denoted by $\bm{T_{logload}}$. Similarly, the time is averaged over all workers that load a log, and over all supersteps during recovery.
\end{itemize}

\begin{table}[!t]
    \centering
	\caption{Time of Checkpointing and Logging}\label{cp_pr}
    \includegraphics[width=0.9\columnwidth]{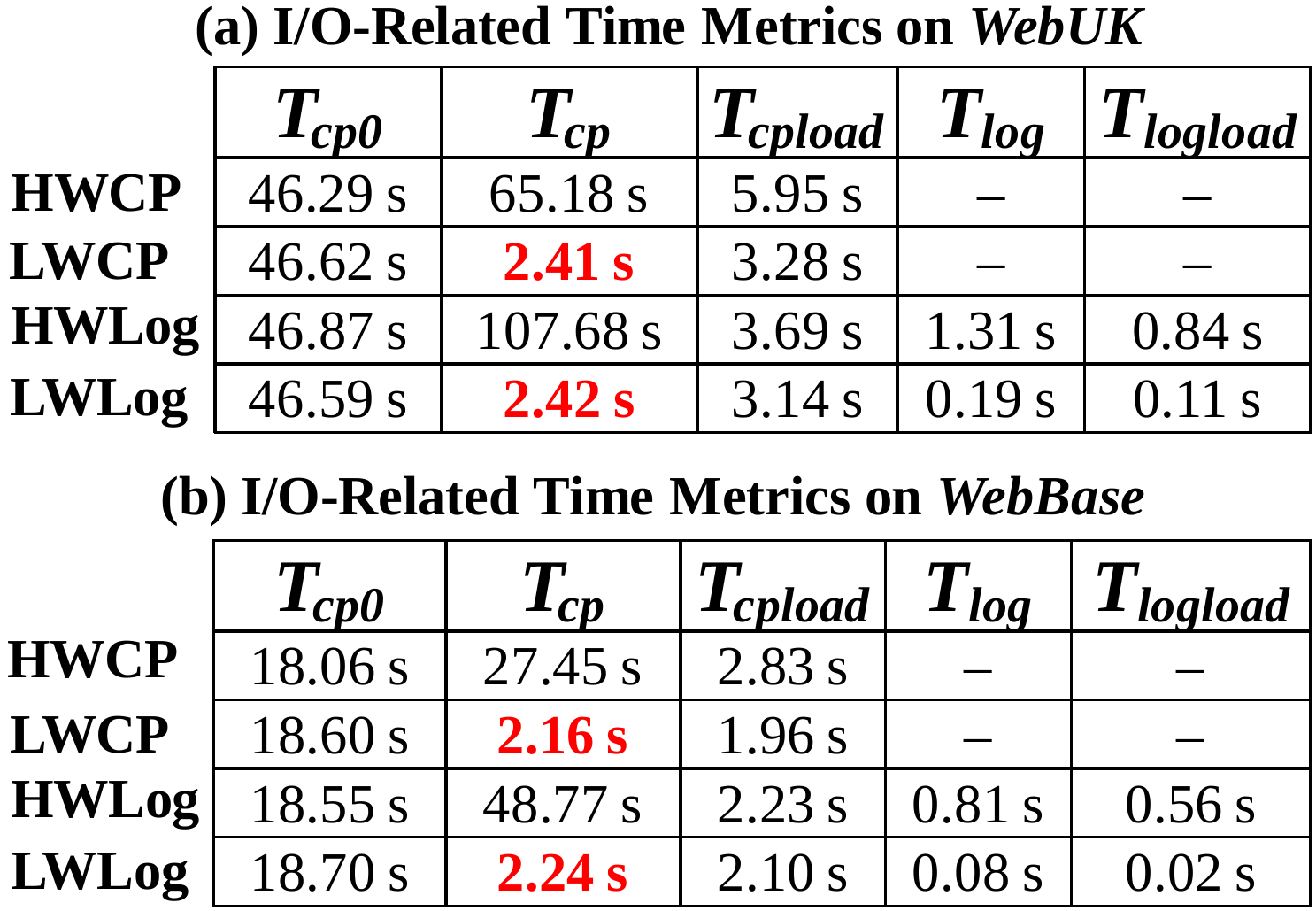}
\end{table}

Table~\ref{cp_pr} reports the above metrics for our PageRank experiments. We can see from Columns with header $T_{cp0}$ that it takes around 46 seconds (resp.\ 18 seconds) to write $CP[0]$ on {\em WebUK} (resp.\ on {\em WebBase}). The time is insensitive to the algorithm adopted, since the content of $CP[0]$ is the same, i.e., vertex states plus adjacency lists.

In contrast, $T_{cp}$ is sensitive to the algorithm adopted. In LWCP and LWLog, $T_{cp}$ is merely less than 2.5 seconds on both datasets, showing that the checkpoints are lightweight. Compared with their corresponding $T_{norm}$ reported in Table~\ref{step_pr}, the checkpointing overhead reported in $T_{cp}$ is negligible. In contrast, in HWCP and HWLog, $T_{cp}$ is a few times that of the corresponding $T_{norm}$ reported in Table~\ref{step_pr}, since the checkpoints written are heavyweight.

Also note that HWLog has a much longer $T_{cp}$ than HWCP. For example, on {\em WebUK} where $T_{norm}$ is around 32 seconds, $T_{cp}$ is 65.18 seconds for HWCP while the time increases to 107.68 seconds for HWLog. This is because HWLog also needs to delete the logged messages for the previous $\delta=10$ supersteps after writing a new checkpoint, which is time consuming due to the large message volume.

Therefore, if garbage collection is performed, HWLog even degrades the failure-free performance compared with HWCP, in return for faster recovery. On the other hand, Table~\ref{cp_pr} shows that LWLog has similar $T_{cp}$ to LWCP, and the additional garbage collection cost of LWLog is negligible. This is because LWLog writes lightweight vertex-state logs (rather than heavyweight message logs as in HWLog).

While we have seen that deleting message logs of $\delta$ supersteps is time-consuming, we find that the cost of log loading/writing is negligible. As Table~\ref{cp_pr} shows, $T_{log}$ is only around 1 second for HWLog, and even much shorter for LWLog. Similarly, $T_{logload}$ is also very short. This is because the OS memory cache provides locality for sequential local reads/writes. Since a worker in our log-based algorithms transmits and logs outgoing messages in parallel and $T_{log}$ is much shorter than $T_{norm}$, logging incurs negligible overhead to normal execution.

\begin{table}[!t]
    \centering
	\caption{Comparison with Other Systems (HWCP only)}\label{compare}
    \includegraphics[width=\columnwidth]{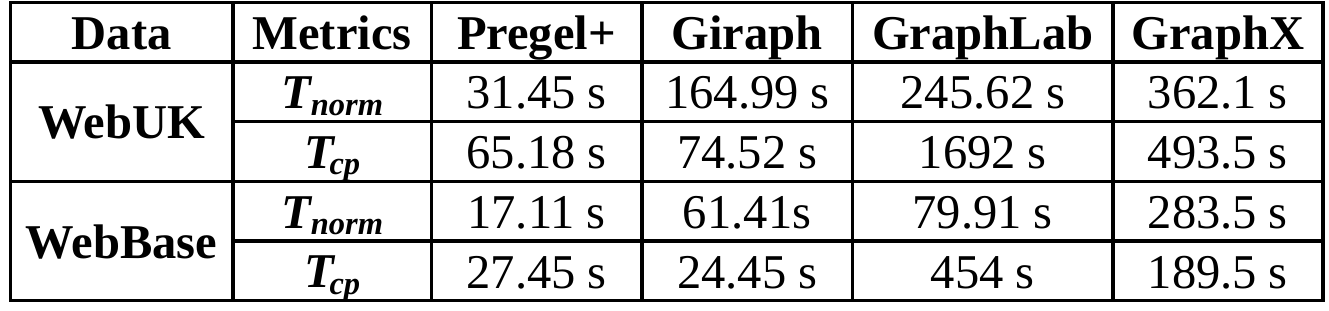}
\end{table}

\begin{table}[!t]
    \centering
    \caption{Performance of HWLog Implementation of [7]}\label{giraph_pr}
	\includegraphics[width=0.86\columnwidth]{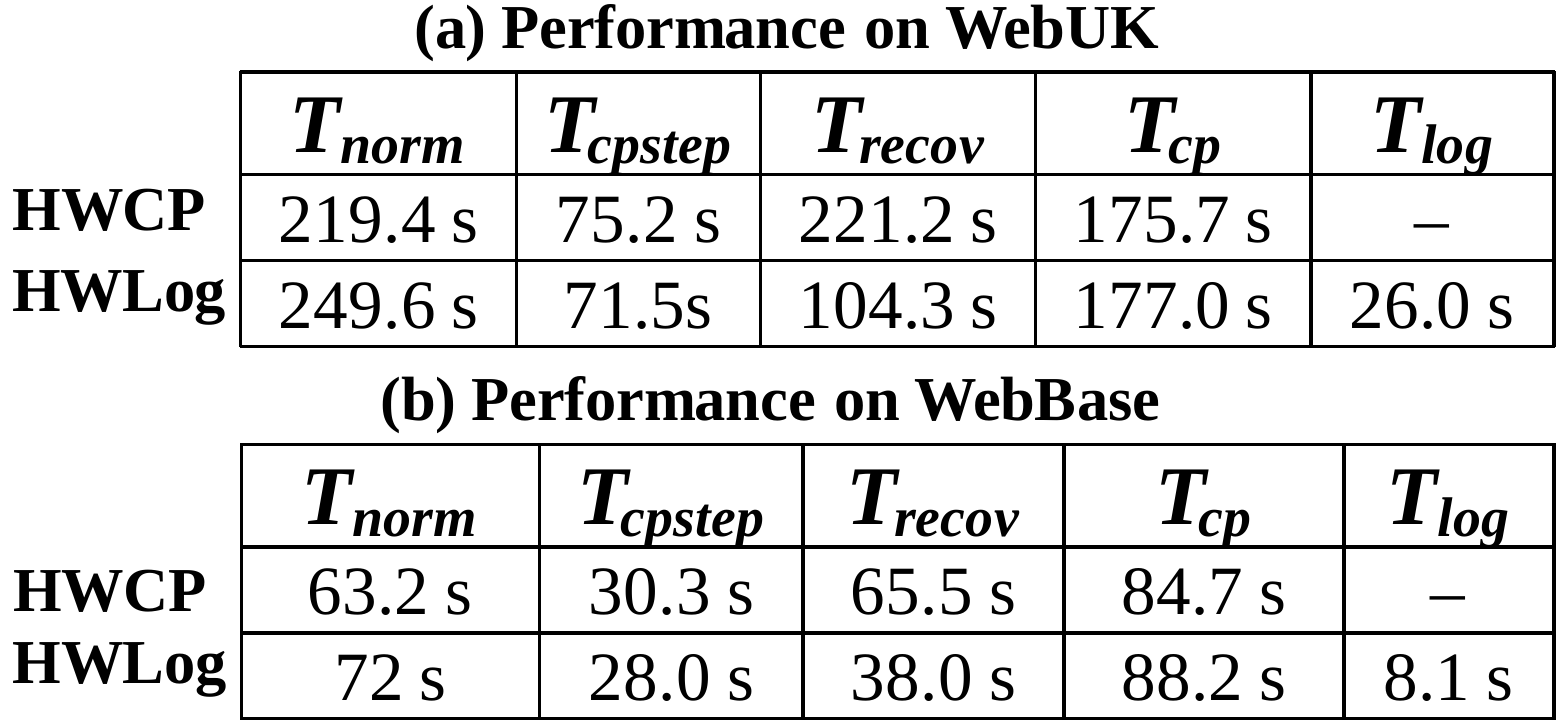}
\end{table}

\vspace{1.5mm}

\noindent{\bf Comparison with Existing Systems.} We have only compared our algorithms on top of our own framework. To show the fairness of the comparison, we now demonstrate that even our baseline algorithm, HWCP, is already faster than existing systems, including Giraph~1.0.0, GraphLab~2.2 and GraphX (Spark~1.1.0), which support only heavyweight checkpointing. For this purpose, we repeated our PageRank experiments on these systems and report the major costs ($T_{norm}$ and $T_{cp}$) in Table~\ref{compare}, which shows that our HWCP implementation has much shorter $T_{norm}$ than the others, and that our $T_{cp}$ is comparable to that of Giraph and much shorter than that of GraphLab and GraphX.

Since~\cite{ftgiraph} implements HWLog in Giraph, we also repeated our PageRank experiments using \cite{ftgiraph}'s system whose code is provided by the authors. Their system does not work properly with the multithreading option of Giraph~1.0.0, and we were only able to run one worker on each machine. Table~\ref{giraph_pr} reports the major costs of their system, which is much higher than our implementation as reported in Tables~\ref{step_pr} and~\ref{cp_pr}.

\subsection{Experiments on Triangle Counting}~\label{exp_tria}
We now report our experiments on triangle counting, whose algorithm is given in the appendix, along with the parameter setting. The performance on both undirected graphs are similar, and thus we only report the experiments on {\em Friendster} and omit those on {\em BTC}. We set checkpointing frequency $\delta=10$ and kill a worker at superstep~20.

Unlike PageRank computation, the time of a superstep decreases with superstep number, and thus average time of a superstep is no longer representative. We redefine the metrics as follows: (1)~$T_{norm}$: the total time taken by running supersteps 11--19 normally before worker failure occurs; (2)~$T_{recov}$: the total time taken by recovering supersteps 11--19 after worker failure is detected; (3)~$T_{cp}$: the time for checkpointing a superstep. We focus only on supersteps between 10 and 20 in order to compare $T_{recov}$ with $T_{norm}$.

Table~\ref{step_tria}(a) shows that log-based algorithms have much smaller $T_{recov}$ than $T_{norm}$, while algorithms writing LWCPs have much smaller $T_{cp}$ than those writing HWCPs. Table~\ref{step_tria}(b) reports $T_{recov}$ when more workers are killed, and an obvious increase in $T_{recov}$ can be observed as the number of failed workers increases.

\begin{table}[!t]
    \centering
    \caption{Triangle Counting Performance on Friendster}\label{step_tria}
	\includegraphics[width=0.7\columnwidth]{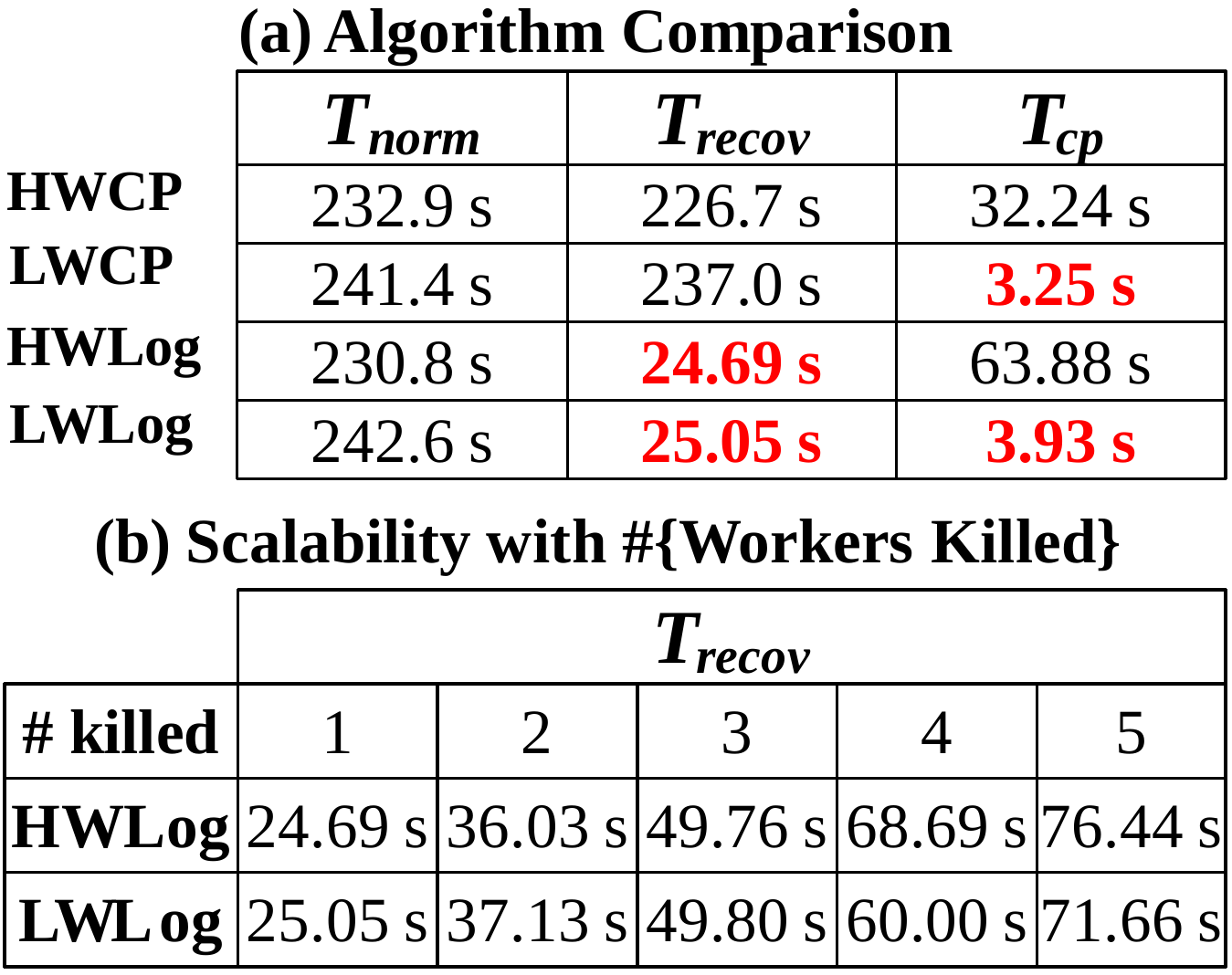}
\end{table}

\section{Conclusions}\label{sec:conclude}
This paper proposed a lightweight checkpointing method that significantly reduces the checkpointing time, and handles challenges like graph mutation and supersteps where LWCP is inapplicable. The idea is further combined with vertex-state log based recovery to reduce recovery time, without sacrificing the benefit of faster checkpointing provided by LWCP. Open-source implementations of our algorithms are provided.


\bibliographystyle{IEEEtran}
\bibliography{IEEEabrv,ref_ftpregel}

\begin{thebibliography}{10}
\providecommand{\url}[1]{#1}
\csname url@samestyle\endcsname
\providecommand{\newblock}{\relax}
\providecommand{\bibinfo}[2]{#2}
\providecommand{\BIBentrySTDinterwordspacing}{\spaceskip=0pt\relax}
\providecommand{\BIBentryALTinterwordstretchfactor}{4}
\providecommand{\BIBentryALTinterwordspacing}{\spaceskip=\fontdimen2\font plus
\BIBentryALTinterwordstretchfactor\fontdimen3\font minus
  \fontdimen4\font\relax}
\providecommand{\BIBforeignlanguage}[2]{{%
\expandafter\ifx\csname l@#1\endcsname\relax
\typeout{** WARNING: IEEEtran.bst: No hyphenation pattern has been}%
\typeout{** loaded for the language `#1'. Using the pattern for}%
\typeout{** the default language instead.}%
\else
\language=\csname l@#1\endcsname
\fi
#2}}
\providecommand{\BIBdecl}{\relax}
\BIBdecl

\bibitem{giraph}
A.~Ching, S.~Edunov, M.~Kabiljo, D.~Logothetis, and S.~Muthukrishnan, ``One
  trillion edges: Graph processing at facebook-scale,'' \emph{{PVLDB}}, vol.~8,
  no.~12, pp. 1804--1815, 2015.

\bibitem{graphlab}
Y.~Low, J.~Gonzalez, A.~Kyrola, D.~Bickson, C.~Guestrin, and J.~M. Hellerstein,
  ``{Distributed GraphLab}: {A} framework for machine learning in the cloud,''
  \emph{{PVLDB}}, vol.~5, no.~8, pp. 716--727, 2012.

\bibitem{powergraph}
J.~E. Gonzalez, Y.~Low, H.~Gu, D.~Bickson, and C.~Guestrin, ``Powergraph:
  Distributed graph-parallel computation on natural graphs,'' in \emph{{OSDI}},
  2012, pp. 17--30.

\bibitem{SalihogluW13ssdbm}
S.~Salihoglu and J.~Widom, ``{GPS:} a graph processing system,'' in
  \emph{{SSDBM}}, 2013, pp. 22:1--22:12.

\bibitem{pregelplus}
D.~Yan, J.~Cheng, Y.~Lu, and W.~Ng, ``Effective techniques for message
  reduction and load balancing in distributed graph computation,'' in
  \emph{{WWW}}, 2015, pp. 1307--1317.

\bibitem{pregel}
G.~Malewicz, M.~H. Austern, A.~J.~C. Bik, J.~C. Dehnert, I.~Horn, N.~Leiser,
  and G.~Czajkowski, ``Pregel: a system for large-scale graph processing,'' in
  \emph{SIGMOD Conference}, 2010, pp. 135--146.

\bibitem{ftgiraph}
Y.~Shen, G.~Chen, H.~V. Jagadish, W.~Lu, B.~C. Ooi, and B.~M. Tudor, ``Fast
  failure recovery in distributed graph processing systems,'' \emph{{PVLDB}},
  vol.~8, no.~4, pp. 437--448, 2014.

\bibitem{survey}
E.~N.~M. Elnozahy, L.~Alvisi, Y.-M. Wang, and D.~B. Johnson, ``A survey of
  rollback-recovery protocols in message-passing systems,'' \emph{ACM Comput.
  Surv.}, vol.~34, no.~3, pp. 375--408, Sep. 2002.

\bibitem{blogel}
D.~Yan, J.~Cheng, Y.~Lu, and W.~Ng, ``Blogel: {A} block-centric framework for
  distributed computation on real-world graphs,'' \emph{{PVLDB}}, vol.~7,
  no.~14, pp. 1981--1992, 2014.

\bibitem{exp}
Y.~Lu, J.~Cheng, D.~Yan, and H.~Wu, ``Large-scale distributed graph computing
  systems: An experimental evaluation,'' \emph{{PVLDB}}, vol.~8, no.~3, pp.
  281--292, 2014.

\bibitem{ppa}
D.~Yan, J.~Cheng, K.~Xing, Y.~Lu, W.~Ng, and Y.~Bu, ``Pregel algorithms for
  graph connectivity problems with performance guarantees,'' \emph{{PVLDB}},
  vol.~7, no.~14, pp. 1821--1832, 2014.

\bibitem{tamerExp}
M.~Han, K.~Daudjee, K.~Ammar, M.~T. {\"{O}}zsu, X.~Wang, and T.~Jin, ``An
  experimental comparison of {Pregel}-like graph processing systems,''
  \emph{{PVLDB}}, vol.~7, no.~12, pp. 1047--1058, 2014.

\bibitem{distsnap}
K.~M. Chandy and L.~Lamport, ``Distributed snapshots: Determining global states
  of distributed systems,'' \emph{{ACM} Trans. Comput. Syst.}, vol.~3, no.~1,
  pp. 63--75, 1985.

\bibitem{optimistic}
S.~Schelter, S.~Ewen, K.~Tzoumas, and V.~Markl, ````{A}ll roads lead to rome'':
  optimistic recovery for distributed iterative data processing,'' in
  \emph{{CIKM}}, 2013, pp. 1919--1928.

\bibitem{replication}
P.~Wang, K.~Zhang, R.~Chen, H.~Chen, and H.~Guan, ``Replication-based
  fault-tolerance for large-scale graph processing,'' in \emph{{DSN}}, 2014,
  pp. 562--573.

\bibitem{ulfm}
W.~Bland, G.~Bosilca, A.~Bouteiller, T.~Herault, and J.~Dongarra, ``A proposal
  for user-level failure mitigation in the mpi-3 standard,'' \emph{Dept. of
  EECS, University of Tennessee}, 2012.

\bibitem{socialnet}
L.~Quick, P.~Wilkinson, and D.~Hardcastle, ``Using pregel-like large scale
  graph processing frameworks for social network analysis,'' in
  \emph{{ASONAM}}, 2012, pp. 457--463.

\bibitem{graphchi}
A.~Kyrola, G.~E. Blelloch, and C.~Guestrin, ``{GraphChi}: Large-scale graph
  computation on just a {PC},'' in \emph{{OSDI}}, 2012, pp. 31--46.

\bibitem{catchw}
Z.~Shang and J.~X. Yu, ``Catch the wind: Graph workload balancing on cloud,''
  in \emph{{ICDE}}, 2013, pp. 553--564.

\bibitem{gpsopt}
S.~Salihoglu and J.~Widom, ``Optimizing graph algorithms on pregel-like
  systems,'' \emph{{PVLDB}}, vol.~7, no.~7, pp. 577--588, 2014.

\bibitem{giraph++}
Y.~Tian, A.~Balmin, S.~A. Corsten, S.~Tatikonda, and J.~McPherson, ``From
  ''think like a vertex'' to ``think like a graph'','' \emph{{PVLDB}}, vol.~7,
  no.~3, pp. 193--204, 2013.

\bibitem{grace}
W.~Xie, G.~Wang, D.~Bindel, A.~J. Demers, and J.~Gehrke, ``Fast iterative graph
  computation with block updates,'' \emph{{PVLDB}}, vol.~6, no.~14, pp.
  2014--2025, 2013.

\bibitem{triangle}
T.~Schank, ``Algorithmic aspects of triangle-based network analysis,''
  \emph{Phd in computer science, University Karlsruhe}, 2007.

\bibitem{triaMapred}
H.~Park, F.~Silvestri, U.~Kang, and R.~Pagh, ``Mapreduce triangle enumeration
  with guarantees,'' in \emph{{CIKM}}, 2014, pp. 1739--1748.

\end{thebibliography}

\appendix
\noindent {\bf LWCP Algorithm for Triangle Finding.} \ The triangle finding algorithm of~\cite{socialnet} consists of two supersteps (assuming that $v1<v2<v3$): (1)~each vertex $v_1$ sends a request $\langle v_1, v_3\rangle$ to its neighbor $v_2$, for all $v_2,v_3 \in \Gamma(v_1)$; (2)~when $v_2$ receives the message, it checks whether $v_3\in\Gamma(v_2)$, and if so, appends $\triangle v_1v_2v_3$ to a file written by its worker on HDFS. We consider a variation where $v_2$ increments its counter in $a(v_2)$ instead, which becomes an algorithm for triangle counting (the total count can be aggregated from the counters of all vertices at last).

This algorithm is vulnerable to failures, since superstep~1 sends at least one request for each triangle, leading to totally $\Omega(|E|^{1.5})$ requests and thus a long-running superstep. We reformulate it into a multi-round variation: in an odd superstep, $v_1$ only sends messages for some $(v_2, v_3)$ pairs, whose number is bounded by $C\cdot|\Gamma(v_1)|$ where $C$ is user-specified; in an even superstep, $v_2$ processes these pairs and increments its counter. This is repeated until there are no more pairs to check for every vertex $v_1$. Thus, the number of messages sent in a superstep is bounded by $C\cdot|V|$.

To implement the algorithm with LWCP, for each vertex $v_1$, we need to include the iterators for outer-loop on $v2$ and inner-loop on $v_3$ into $a(v_1)$, so that $v_1$ can continue to iterate for more pairs $(v_2, v_3)$ in each round. A pitfall here is to implement it as in HWCP where the iterating direction in {\em compute}(.) is always forward, which is incorrect with LWCP as we explain below. To generate messages for superstep~$i$, we iterate from $a^{(i-1)}(v_1)$ to $a^{(i)}(v_1)$ during normal execution; but when we generate messages from the state $state^{(i)}(v_1)$ that is loaded from $CP[i]$, we should reverse iterate the iterators from $a^{(i)}(v_1)$ back to $a^{(i-1)}(v_1)$ to generate the same set of messages, and iterating forward results in incorrect messages.

Thus, we follow Equations~(\ref{eq:g}) and~(\ref{eq:h}) exactly when writing $v_1$.{\em compute}(.). We first iterate forward for at most $C\cdot|\Gamma(v_1)|$ pairs, but we only update the iterators in $a(v_1)$ without generating messages. Then, we reverse iterate from the updated iterators back to generate messages. This implementation works correctly with LWCP.

We report the performance of this algorithm in Section~\ref{exp_tria}. Unlike in PageRank, the time of each round decreases as the algorithm runs on, since more and more vertices exhaust their neighbor-pairs. Therefore, the algorithm is more suitable for time-interval based checkpointing, although we performed checkpointing every 10 supersteps for the convenience of running experiments. We set $C=1$ in our experiments on {\em Friendster}, since {\em Friendster} has a high average degree, and the total message volume in superstep~1 already exceeds the memory size of our cluster when $C>3$.

\end{document}